Unconventional satellite resistance peaks in moiré superlattice of *h*-BN/ AB-stacked tetralayer-graphene heterostructure

Fumiya Mukai[1], Kota Horii[1], Ryoya Ebisuoka[2], Kenji Watanabe[2], Takashi Taniguchi[3] and Ryuta Yagi[1*]

[1]Graduate School of Advanced Sciences and Engineering, Hiroshima University,
1-3-1 Kagamiyama, Higashi-Hiroshima 739-8530, Japan,
[2]Research Center for Functional Materials, National Institute for Materials Science, 1-1 Namiki, Tsukuba 305-0044, Japan,
[3]International Center for Materials Nanoarchitectonics, National Institute for Materials Science, 1-1 Namiki, Tsukuba 305-0044, Japan.

ABSTRACT

Most studies on moiré superlattices formed from a stack of *h*-BN and graphene have focused on single layer graphene; graphene with multiple layers is less understood. Here, we show that a moiré superlattice of multilayer graphene shows new features arising from the anisotropic Fermi surface affected by the superlattice structure. The moiré superlattice of a *h*-BN/AB-stacked tetralayer graphene heterostructure exhibited resistivity peaks showing a complicated dependence on the perpendicular electric field. The peaks were not due to secondary Dirac cones forming, but rather opening of the energy gap due to folding of the anisotropic Fermi surface. In addition, superlattice peaks resulted from mixing of light- and heavy-mass bilayer-like bands via the superlattice potential. The gaps did not open on the boundary of the superlattice Brillouin zone, but rather opened inside it, which reflected the anisotropy of the Fermi surface of multilayer graphene.

*Corresponding author: yagi@hiroshima-u.ac.jp

Introduction

Moiré is a kind of geometrical interference that appears when two nearly identical lattices are aligned with a small angle mismatch (Fig. 1(a)). The discovery of moiré in a stack of *h*-BN and graphene [1-2] has led to numerous studies associated with superstructures of graphene formed by incommensurate substrates [3-12]. The moiré structure virtually gives graphene another periodic structure, and therefore, these systems can be regarded as superlattices [13-14]. The conventional understanding of superlattices is that an energy gap forms at the boundary of the Brillouin zone of the superlattice. However, recent findings in monolayer graphene have shown that Dirac cones are formed by locally closing the energy gap, and this results in additional peaks in the carrier density dependence of resistance [3-6,10]. Further studies have indicated that the secondary Dirac cones are formed at the K or K' point of the superlattice Brillouin zone [15-18]. The moiré potential forms a band gap of about 20 meV at the secondary Dirac cone in the hole branch [7,12]. In addition, the superlattice potential forms a band gap of about~ 35 meV at the primary Dirac cone in the case of perfect alignment, which appears at the charge neutrality point [4-7,12,19-20]. These band-gap values are highly dependent on the alignment angle and sample structure [4,12].

Although there have been a number of studies on moiré superlattices of *h*-BN/single layer graphene heterostructures, there have only few studies on moiré superlattices of graphene with multiple layers. As far as the authors know, bilayer graphene has been studied in terms of Hofstadter butterfly's diagram in the energy spectrum obtained in a magnetic field [9,18,21]. The superlattice potential due to the moiré would primarily influence the surface layer of graphene in contact with it. Therefore, it is not clear how the electronic band structure is modified by the moiré potential in graphene with multiple layers. Moreover, trigonal warping of the band structure becomes increasingly dominant as the number of layers increases [22-23]. This effect would be important in forming an energy gap due to the moiré structure, and it would also be important in reconstructing the Fermi surface. In this work, we studied the electronic band structure of a moiré superlattice of AB-stacked tetralayer graphene, as an example of multilayer graphene, by conducting low-temperature transport experiments. We found new features intrinsic to the moiré superlattice of multilayer graphene.

The band structure of AB-stacked tetralayer graphene consists of two bilayer-like bands (light and heavy mass) as shown in Fig. 1(b). A simplified band model indicates that the tetralayer graphene should have a semi-metallic band in which there is an overlap of the electron and hole bands near $E = 0$; however, mixing between the heavy and light-mass bilayer-like bands creates local energy gaps, and thus, the band overlap of the conduction and valence bands is expected to be considerably smaller than predicted by the simplified band model [24-27]. The band overlap vanishes by applying a perpendicular electric field; an energy gap opens at the bottoms of the bilayer-like bands [25-26] as in the case of bilayer graphene [28-36]. The bottoms of the heavy-mass and light-mass bilayer-like bands become nearly flat [25-27]. However, trigonal warping locally closes the energy gap, resulting in mini-Dirac cones at the bottom of the heavy-mass bilayer-like band. These characteristic structures in the dispersion relation were revealed by studying the intrinsic resistance peaks that as a function of carrier density and perpendicular electric flux density [25-26,37-38].

Results and Discussion

Figure 1 (c) shows an optical micrograph of an AB-stacked tetralayer graphene sample that consisted of encapsulated graphene with a top and a bottom gate electrode. The low-temperature electrical mobility of this device was above $4 \times 10^4$ cm$^{-2}$/Vs. As shown in Figure 1 (d), peaks which arise from moiré superlattice appear in the plot of resistivity ($\rho$) against bottom gate voltage ($V_b$). In addition to the clear main peak at $V_b \approx 0$ V, a couple of side peaks are discernible at $V_b \approx 40$ and $-45$ V, which are reminiscent of superlattice peaks due to the secondary Dirac points in a moiré superlattice formed with $h$-BN and monolayer (or bilayer) graphene.

The resistivity traces ($\rho$ $vs$ $V_b$) showed significant variation with top gate voltage (Fig. 2(a)), and detailed measurements of the resistivity map revealed a fine structure of resistivity peaks (Fig. 2(b)). The straight ridge from the top left to bottom right of Fig. 2(b) is the condition of charge neutrality, which is given by $n_{tot} = 0$, where $n_{tot}$ is the carrier density induced by the top and bottom gate voltage,

$$n_{tot} = (C_t V_t + C_b V_b)/e. \tag{1}$$

Here, $C_t$ and $C_b$ are the specific capacitances for the top and bottom gate electrodes. The perpendicular electric flux density ($D_\perp$), which is roughly proportional to the strength of the perpendicular electric field, can be calculated as

$$D_\perp = (C_t V_t - C_b V_b)/2. \tag{2}$$

Figure 3(a) is a replot of the map. One can see ridges of resistivity (bb+, bb-, MDP) that were recently found in AB-stacked tetralayer graphene [25-27,37-38]. Ridges bb+ and bb- reflect the bottoms of the bilayer-like bands, whose dispersions are nearly flat in a perpendicular electric field. MDP is due to mini-Dirac cones. However, the other ridges (A+, B+, C+, *etc.*) are absent from pristine AB-stacked tetralayer graphene.

The Landau levels were also influenced by the moiré structure. Figure 1 (e) shows a map of longitudinal resistivity ($\rho_{xx}$), which was measured as a function of $n_{tot}$ and magnetic field ($B$) for $D_\perp = 0$ cm$^{-2}$As. Each streak indicated by bright and dark lines is a Landau level and the energy gap between them. The Landau level structure between $n_{tot} = -2 \times 10^{12}$ and $2 \times 10^{12}$ cm$^{-2}$ are specific to pristine AB-stacked tetralayer graphene [25-27,39]. In addition, at $n_{tot} \sim 3.5 \times 10^{12}$ cm$^{-2}$, one can see a Landau fan reminiscent of that for the secondary Dirac cones in the case of moiré superlattice in bilayer graphene. Around this carrier density, ridge A+ appears (Fig. 2(c)). Such Landau fans could not be clearly seen in the hole regime, which implies that the moiré superlattice for multilayer graphene is different from those in the mono- or bilayer case.

To study the origin of the ridges, we compared the results of the experiment with a theoretical calculation. Figure 2 (d) shows a map of resistivity that was numerically calculated from the dispersion relations by using Boltzmann transport theory with the constant relaxation time approximation. The band structure calculation took account of an effective moiré potential based on the *h*-BN-graphene hopping model presented by Moon and Koshino [18]. The amplitude of the effective potential and mismatch angle of the crystal axes between *h*-BN and graphene were adjusted so that the calculation reproduced the experimental results (see the Supplementary material). The calculation approximately reproduced the experimental map of resistivity, which indicates that the moiré potential significantly influenced the band structure of the multilayer graphene even though the potential is present at the first layer of graphene contacting *h*-BN. The ridges bb+ and bb- are intrinsic to AB-stack tetralayer graphene [25-26, 37-38]. By studying the relation between the shape of the Fermi surface (the energy contour of the dispersion relation) and carrier density, it was found that the other ridges appeared when an energy gap associated with the superlattice potential opened. These resistivity peaks (ridges) are due to formation of an energy gap associated with nesting of the Fermi surfaces which are translated by reciprocal vectors in the extended zone scheme, as shown in Fig 3(a)-(c). Top panels in the figures show the numerically calculated Fermi surface(s) for different energies in the case of $|D_\perp| = 3.2 \times 10^{12}$ cm$^{-2}$As. Bottom panels show illustrations. The Fermi surface of AB-stacked tetralayer graphene consists of trigonally warped circles. It changes its topology by opening an energy gap when the Fermi surfaces are nested in the superlattice potential. Opening an energy gap reduces the electronic states available for electrical conduction, and thereby reduces conductivity. Ridge A+ (A-) are for nesting of the *Be* (*Bh*) band. Ridges B+ (B-) and C+ (C-) are for nesting of *Be* (*Bh*) and *be* (*bh*), which occur at different carrier densities (energies) because of the trigonally warped Fermi surface (Fig. 3 (b) and (c)). For higher carrier densities, nesting *be* (*bh*) with itself would possibly occur and it should result in a corresponding resistance ridge.

Figure 2 (e) shows simplified band diagrams in the extended zone scheme to explain the opening of the energy gap; FBZ is the first Brillouin zone, and identical band diagrams are translated by reciprocal vectors. Small energy gaps open when bands cross each other, and this results in resistance ridges A+, B+, C+, *etc.*; for example, ridge A+ occurs when band *Be* crosses itself in the neighboring zone; ridge B+ occurs when *be* crosses *Be*, *etc.* Because the band structure is trigonally warped ([22-23]), the energy gaps associated with the nesting generally occur at wave numbers other than on the Brillouin zone boundary, as can be seen in the figure.

The complicated dependence of the peaks (ridges) on $n_{tot}$ and $D_\perp$ is due to the multiband property relevant to the variation of dispersion relations through the perpendicular electric field. For example, ridge A+ (A-) bends to smaller $|n_{tot}|$ with increasing $|D_\perp|$. This should occur because the energy gap between *Be* and *be* (*Bh* and *bh*) widens with increasing $|D_\perp|$, as shown in Fig. 2(e). At a chemical potential in the vicinity of the energy gap for peak A+ (A-) (the dashed line in Fig. 2(e)), for example, the number of carriers in *be* decreases with increasing $|D_\perp|$ because the difference in energy between the bottom of *be* and the bottom of *Be* increases with increasing $|D_\perp|$ (Fig. 2(c), center). Above the critical $|D_\perp|$, the energy of the bottom of *be* (*bh*) becomes larger than the chemical potential (dashed line), and only the Fermi surface of *Be* (*Bh*) is present (Fig. 2(c) right). This transition can be seen in the map (Figs. 2(c) and (d)) as crossings of ridge A+ (A-) and bb+ (bb-) at $|n_{tot}| \sim 2.3 \times 10^{12}$cm$^{-2}$ and $|D_\perp| \sim 3.5 \times 10^{-7}$ cm$^{-2}$As. The electron-hole asymmetry in the map can be similarly explained by considering the electron-hole asymmetry in the dispersion relation. The Fermi surface areas of the light-mass bilayer-like band and the heavy-mass one have different ratios in electron and hole regimes because of this asymmetry.

As shown above, the resistance peaks (ridges) due to the moiré superlattice can be explained by energy gaps opening. It is commonly accepted that energy gaps form on the Brillouin zone boundary of the superlattice. Indeed, the energy gap is expected to open at the Brillouin zone boundary in moiré superlattice for monolayer graphene which is approximately isotropic. However, it can open inside the Brillouin zone in an anisotropic electron system (Fermi surface) as in the case of AB-stacked graphene with multiple layers. Textbooks on solid-state physics tell us that, in the presence of a periodic potential, an energy gap opens via interference of wave functions $\phi(\vec{k})$ and $\phi(\vec{k} \pm \vec{G})$, where $\vec{k}$ is a wave vector and $\vec{G}$ is one of the reciprocal vectors. If the original band structure satisfies $E(\vec{k}) = E(\vec{k} \pm \vec{G})$, an energy gap should open on a boundary of the Brillouin zone. On the other hand, if the original band is anisotropic, as in the case of AB-stacked multilayer graphene, the condition, $E(\vec{k}) = E(\vec{k} \pm \vec{G})$, can be satisfied and an energy gap should form even though $\vec{k}$ is not on the boundary of but rather inside the Brillouin zone, as can be seen in Fig. 3(a). Similar phenomena arising from anisotropy have recently been discussed in photonic crystals [40] and phonic crystals [41]; Bragg reflection occurs inside the Brillouin zone in anisotropic media, rather than on the boundary. In addition, an energy gap opens when the Fermi surfaces of different bands are nested. The wave functions in band 1, $\phi_1(\vec{k_1})$, and in band 2, $\phi_2(\vec{k_2})$, which are orthogonal to each other in the absence of moiré potential, anti-cross when $E(\vec{k_1}) = E(\vec{k_2} \pm \vec{G})$ is satisfied. This can be seen in Fig. 3 (a) and (c). Energy gaps open and the Fermi surfaces change their topology when *be* and *BE* are nested.

The detailed structure of the numerically calculated dispersion relation is highly dependent on the models of the effective moiré potential and its amplitude. However, the essential feature of the Fermi surface topology is unchanged by the choice of model. We calculated the dispersion relations and resistivity maps for *h*-BN-graphene hopping models [16-18,42], the 2D charge modulation model [43], and the potential modulation model [3,16,42] (see the Supplementary material). The *h*-BN-graphene hopping models approximately reproduce the experimental resistivity map if the potential amplitude is less than about half of those given in Refs. [16-18,42]. Using any of these models, the resistivity map was approximately reproduced for a sufficiently small potential amplitude, which indicates that the topological transitions of the Fermi surface due to nesting occur regardless of the model used.

If the resistivity peaks (ridges) originate from nesting of the Fermi surface, the resistivity maps should show a significant dependence of the trigonal warping because the nesting is significantly influenced by the shape of the Fermi surface. Figure 4 shows the maps calculated for different values of $\gamma_3$ of the SWMcC parameters, which tunes the trigonal warping; $\gamma_3 = 0$ eV is for circular Fermi surfaces; $\gamma_3 = 0.3$ eV approximates the experiment. Nesting of *Be* and *be* (*Bh* and *bh*) can occur at different points, and thereby, at different energies (or carrier densities) in the trigonally warped case as illustrated in Figs. 3(b) and (c). This results in separate ridges B+ (B-) and C+ (C-) ($\gamma_3 = 0.3$ eV in Fig. 4). With decreasing $\gamma_3$, ridges B+ and C+ (B- and C-) merge into a single ridge ($\gamma_3 = 0$ eV in Fig. 4). In this case, the energy gap should open on the boundary of the superlattice Brillouin zone.

Finally, we comment on the period of the superlattice potential and mismatch angle between the crystallographic axes of *h*-BN and that of graphene. In the case of the moiré superlattice of *h*-BN/monolayer graphene, one can calculate the period of the moiré potential [3-6] from the size of the superlattice Brillouin zone, which can be estimated from the carrier density of the resistivity peak due to the superlattice ($n_{FBZ}$) [3-6]. The mismatch angle $\phi$ can be estimated as [3-6],

$$n_{FBZ} = \frac{8}{\sqrt{3}\lambda^2} \qquad (3)$$

where

$$\lambda = \frac{(1+\delta)a}{\sqrt{2(1+\delta)(1-\cos\phi)+\delta^2}} \ . \qquad (4)$$

Here, $\lambda$ is the wave length (or lattice constant) of the moiré potential, $\delta$ is the difference between the lattice constants of *h*-BN and graphene, and $a$ is the lattice constant of graphene [3]. In the multiband system, however, $n_{fBZ}$ does not reflect the size of the superlattice Brillouin zone. For this reason, in the present experiment, the mismatch angle (and the period of the superlattice) was determined by numerically calculating the maps of resistivity for different angles and adjusting them so that the calculation reproduced the experimental map (see the Supplementary material). The experimental map was approximately reproduced for $\theta = 0.35°$. The period of the moiré potential was about 53 times the lattice constant of graphene.

### Summary and Concluding remarks

The moiré superlattice of a *h*-BN/AB-stacked tetralayer graphene heterostructure was studied by conducting low-temperature transport measurements. It was found that the carrier density dependence of the resistivity showed superlattice peaks reminiscent of those in the monolayer case. The electronic band structure was strongly influenced by the moiré potential which would act only the outermost layer of graphene that contacts *h*-BN. A multiband property was observed in the significant dependence of the superlattice peaks on the perpendicular electric field. These peaks were significantly influenced by the trigonal warping of the band structure. The peaks occurred when the Fermi surfaces translated through a reciprocal vector are nested, which resulted in the energy gap opening and a topological transition of the Fermi surface. The energy gap opens inside the superlattice Brillouin zone rather than on its boundary. These are new aspects of the moiré superlattice of the *h*-BN/graphene heterostructure and two-dimensional superlattices with anisotropic band structures.

**Acknowledgements**

This work was supported in part by KAKENHI No. 25107003 from MEXT Japan. K.W. and T.T. acknowledge support from the Elemental Strategy Initiative conducted by the MEXT, Japan ,Grant Number JPMXP0112101001,  JSPS KAKENHI Grant Number JP20H00354 and the CREST(JPMJCR15F3), JST.

Figure captions

**Fig. 1 Sample Characterization**

(a) Schematic drawing of moiré structure formed by stacking two honeycomb lattice sheets. The arrows indicate the periodic structure. (b) Low energy dispersion relation of AB-stacked tetralayer graphene in the absence (left) and presence (right) of a perpendicular electric flux density, calculated for Slonczewski-Weiss-McClure (SWMcC) parameters of graphite. $k_0 = \frac{2\gamma_1}{\sqrt{3}\gamma_0 a}$ ($a$ is the lattice constant of graphene). FLB stands for the nearly flat part in the band. (c) Optical micrograph of a sample. Hall bar equipped with top and bottom gate electrodes. The bar is 10 μm long. (d) Bottom gate voltage ($V_b$) dependence of resistivity ($\rho$) measured for zero top gate voltage ($V_t$ = 0 V). $T$ = 4.2 K and $B$ = 0 T. The peak for charge neutrality ($V_b \approx$ 0 V) and satellite peaks ($V_b \approx$ 40 and $-45$ V) are discernible. (d) Map of longitudinal resistivity $\rho_{xx}(B)$ vs $n_{tot}$ (total carrier density). $T$ = 4.2 K. Perpendicular electric flux density ($D_\perp$) was kept at zero by tuning $V_b$ and $V_t$. The arrow indicates Landau levels that are reminiscent of those of the secondary Dirac cone in the moiré superlattice of the h-BN/ single layer graphene heterostructure. (e) Map of longitudinal resistivity ($\rho_{xx}$) measured for $D_\perp = 0$.

**Fig. 2 Resistance peaks due to moiré superlattice**

(a) $V_b$-dependence of $\rho$ for different $V_t$. ($T$ = 4.2 K). (b) Map of $\rho$ plotted as a function of $V_b$ and $V_t$. ($T$ = 4.2 K). (c) Map of $\rho$ plotted as a function of carrier density ($n_{tot}$) and perpendicular electric flux density ($D_\perp$). Some of the salient ridges are labeled as A+, A-, B+, B-, C+, C-, bb+, bb-, and MDP. $T$ = 4.2 K and $B$ = 0 T. (d) Map of $\rho$ numerically calculated by using the effective mass approximation with the SWMcC parameters of graphite. The effective moiré potential given in Ref. [18] was used with a mismatch angle of 0.35°. The amplitude of the moiré potential was scaled with a factor of 0.5. Labels correspond to those in panel (c). Ridges BB+ and BB-, which are not clearly visible in panel (c), arose from the bottom of the heavy-mass bilayer-like band. (e) Schematic illustration of dispersion relations for different $|D_\perp|$ in the extended zone scheme. *Be, Bh, be,* and *bh* are bilayer-like bands in AB-stacked tetralayer graphene. Energy gaps open at the points of band crossings where Fermi surfaces separated by a reciprocal vector are nested (shown by the open circles). This results in resistivity peaks when the chemical potential is swept. At the chemical potential where A+ appears (dashed line), there are two electron bands for $D_\perp = 0$. With increasing $|D_\perp|$, the difference in energy between the bottoms of *Be* and *be* increases. Accordingly, the carrier density decreases, as shown by the size of the Fermi surface of be (FSbe) in the middle panel. The Fermi surface of *be* disappears at the critical $|D_\perp|$ (right).

**Fig. 3  Topological transition of Fermi surface**

Example energy contours of the numerically calculated dispersion relation for different $n_{tot}$. $|D_\perp| = 3.2 \times 10^{-7}$ cm$^{-2}$As (top panels). Bottom panels are schematic illustrations. Hexagon XYXYXY is the first Brillouin zone of the superlattice potential. $G_1$-$G_6$ are reciprocal vectors of the superlattice. For a weak potential, the energy gap does not open on the Brillouin zone boundary. The energy gap opens at the point where the Fermi surfaces in a different Brillouin zone are nested (shown by circles). This results in a topological transition of the Fermi surface. Panel (a) shows the case of nesting of *Be* with itself. Panels (b) and (c) show different cases of nesting of *be* and *Be*.

**Fig. 4  Effect of trigonal warping on superlattice peaks.**

Numerically calculated maps of $\rho$. From left to right, $\gamma_3$ is 0, 0.1, and 0.3 eV. The rightmost panel shows the experimental results. Some of the characteristic ridges are labeled. Ridges B+ (B-) and C+ (C-) merge into a single ridge BC+ (BC-). This indicates that the superlattice peaks are strongly dependent of the shape (or symmetry) of the Fermi surface.

[1] Decker, R. *et al.*, Local Electronic Properties of Graphene on a BN Substrate via Scanning Tunneling Microscopy, Nano Lett. **11**, 2291 - 2295 (2011).

[2] Xue, J. *et al.*, Scanning tunnelling microscopy and spectroscopy of ultra-flat graphene on hexagonal boron nitride, Nat. Mater. **10**, 282 - 285 (2011).

[3] Yankowitz, M. *et al.*, Emergence of superlattice Dirac points in graphene on hexagonal boron nitride, Nat. Phys. **8**, 382 - 386 (2012).

[4] Hunt, B. *et al.*, Massive Dirac Fermions and Hofstadter Butterfly in a van der Waals Heterostructure, Science **340**, 1427 - 1430 (2013).

[5] Woods, C. *et al.*, Commensurate-incommensurate transition in graphene on hexagonal boron nitride, Nat. Phys. **10**, 451 - 456 (2014).

[6] Ponomarenko, L. *et al.*, Cloning of Dirac fermions in graphene superlattices, Nature **497**, 594 - 597 (2013).

[7] Yankowitz, M. *et al.*, Dynamic band-structure tuning of graphene moire superlattices with pressure, Nature **557**, 404 (2018).

[8] Chen, G. *et al.*, Emergence of Tertiary Dirac Points in Graphene Moire Superlattices, Nano Lett. **17**, 3576 - 3581 (2017).

[9] Dean, C. *et al.*, Hofstadter's butterfly and the fractal quantum Hall effect in moire superlattices, Nature **497**, 598 - 602 (2013).

[10] Yu, G. *et al.*, Hierarchy of Hofstadter states and replica quantum Hall ferromagnetism in graphene superlattices, Nat. Phys. **10**, 525 - 529 (2014).

[11] Pezzini, S. *et al.*, Field-induced insulating states in a graphene superlattice, Phys. Rev. B **99**, 045440 (2019).

[12] Wang, L. *et al.*, Evidence for a fractional fractal quantum Hall effect in graphene superlattices, SCIENCE **350**, 1231 - 1234 (2015).

[13] Esaki,L. & Tsu,R., Superlattice and negative differential conductivity in semiconductors, IBM J. Research. and Development. **14**, 61 (1970).

[14] Chang, L. L., Sakaki, H., Chang, C. A., & Esaki, L., Phys. Rev. Lett., **38,** 1489-1493, (1977) .

[15] Shi, Z. *et al.*, Gate-dependent pseudospin mixing in graphene/boron nitride moire superlattices, Nat. Phys. **10**, 743 - 747 (2014).

[16] Wallbank, J., Patel, A., Mucha-Kruczynski, M., Geim, A. & Falko, V., Generic miniband structure of graphene on a hexagonal substrate, Phys. Rev. B **87**, 245408 (2013).

[17] Kindermann, M., Uchoa, B. & Miller, D., Zero-energy modes and gate-tunable gap in graphene on hexagonal boron nitride, Phys. Rev. B **86**, 115415 (2012).

[18] Moon, P. & Koshino, M., Electronic properties of graphene/hexagonal-boron-nitride moire superlattice, Phys. Rev. B **90**, 155406 (2014).

[19] Chen, Z. *et al.*, Observation of an intrinsic bandgap and Landau level renormalization in graphene/boron-nitride heterostructures, Nat. Commun. **5**, 4461 (2014).

[20] Kim, H. *et al.*, Accurate Gap Determination in Monolayer and Bilayer Graphene/h-BN Moire Superlattices, Nano Lett. **18**, 7732 - 7741 (2018).

[21] Mucha-Kruczynski, M., Wallbank, J. & Fal'ko, V., Heterostructures of bilayer graphene and h-BN, Phys. Rev. B **88**, 205418 (2013).

[22] Oka, T. *et al.*, Ballistic Transport Experiment Detects Fermi Surface Anisotropy of Graphene, Phys. Rev. B **99**, 035440 (2019).

[23] Tajima, S., Ebisuoka, R., Watanabe, K., Taniguchi, T. & Yagi, R., Multiband Ballistic Transport and Anisotropic Commensurability Magnetoresistance in Antidot Lattices of AB-stacked Trilayer Graphene, J. Phys. Soc. Jpn. **89**, 044703 (2020).

[24] Koshino, M. & McCann, E., Landau level spectra and the quantum Hall effect of multilayer graphene, Phys. Rev. B **83**, 165443 (2011).

[25] Shi, Y.M. *et al.*, Tunable Lifshitz Transitions and Multiband Transport in Tetralayer Graphene, Phys. Rev. Lett. **120**, 096802 (2018).

[26] Hirahara, T. *et al.*, Multilayer Graphene Shows Intrinsic Resistance Peaks in The Carrier Density Dependence, Sci. Rep. **8**, 13992 (2018).

[27] Horii, K. *et al.*, Magnetotransport study of the mini-Dirac cone in AB-stacked four- to six-layer graphene under perpendicular electric field, Phys. Rev. B **100**, 245420 (2019).

[28] Castro, E. V. *et al.*, Biased bilayer graphene, Phys. Rev. Lett. **99**, 216802 (2007).

[29] Avetisyan, A. A., Partoens, B. & Peeters, F. M., Electric-field control of the band gap and Fermi energy in graphene multilayers by top and back gates, Phys. Rev. B **80**, 195401 (2009).

[30] Avetisyan, A. A., Partoens, B. & Peeters, F. M., Electric field tuning of the band gap in graphene multilayers, Phys. Rev. B **79**, 035421 (2009).

[31] Koshino, M. & McCann, E., Gate-induced interlayer asymmetry in ABA-stacked trilayer graphene, Phys. Rev. B **79**, 125443 (2009).

[32] Min, H. K., Sahu, B., Banerjee, S. K. & MacDonald, A. H., Ab Initio theory of gate induced gaps in graphene bilayers, Phys. Rev. B **75**, 155115 (2007).

[33] Taychatanapat,T. & Jarillo-Herrero,P., Electronic Transport in Dual-gated Bilayer Graphene at Large Displacement Fields, Phys. Rev. Lett. **105**, 166601 (2010).

[34] Miyazaki,H., Tsukagoshi,K., Kanda,A., Otani,M. & Okada,S., Influence of Disorder On Conductance in Bilayer Graphene Under Perpendicular Electric Field, Nano Lett. **10**, 3888-3892 (2010).

[35] Yan,J. & Fuhrer, M.S., Charge Transport in Dual Gated Bilayer Graphene With Corbino Geometry, Nano Lett. **10**, 4521-4525 (2010).

[36] Ohta, T., Bostwick, A., Seyller, T., Horn, K. & Rotenberg, E., Controlling the electronic

structure of bilayer graphene, Science **313**, 951-954 (2006).

[37] Nakasuga, T. *et al.*, Intrinsic resistance peaks in AB-stacked multilayer graphene with odd number of layers, Phys. Rev. B **101**, 035419 (2020).

[38] Nakasuga, T. *et al.*, Low-energy Band Structure in Bernal Stacked Six-layer Graphene, Phys. Rev. B **99**, 085404 (2019).

[39] Yagi, R. *et al.*, Low-energy band structure and even-odd layer number effect in AB-stacked multilayer graphene, Sci. Rep. **8**, 13018 ( 2018).

[40] Sivarajah, P., Maznev, A., Ofori-Okai, B. & Nelson, K., What is the Brillouin zone of an anisotropic photonic crystal?, Phys. Rev. B **93**, 054204 (2016).

[41] Wang, Y., Maznev, A. & Laude, V., Formation of Bragg Band Gaps in Anisotropic Phononic Crystals Analyzed With the Empty Lattice Model, CRYSTALS **6**, 52 (2016).

[42] Wallbank, J., Mucha-Kruczynski, M., Chen, X. & Fal'ko, V., Moire superlattice effects in graphene/boron-nitride van der Waals heterostructures, Ann. Phys. **527**, 359 - 376 (2015).

[43] Ortix, C., Yang, L. & van den Brink, J., Graphene on incommensurate substrates, Phys. Rev. B **86**, 081405 (2012).

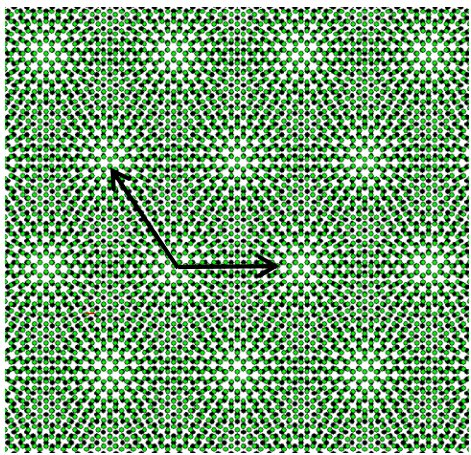
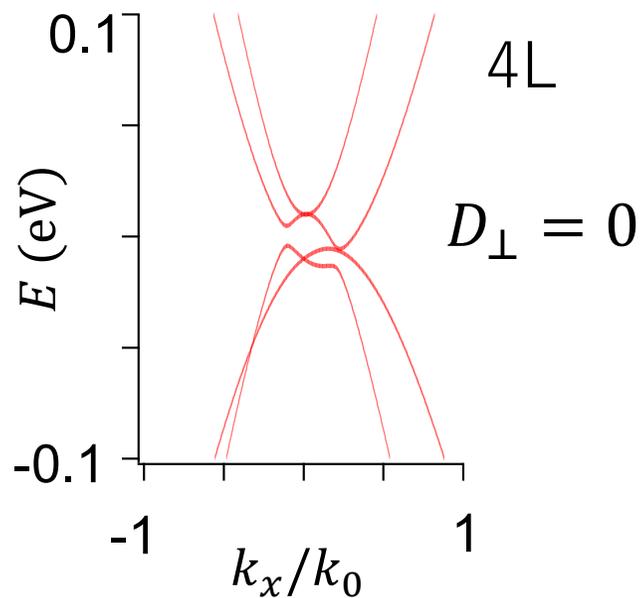
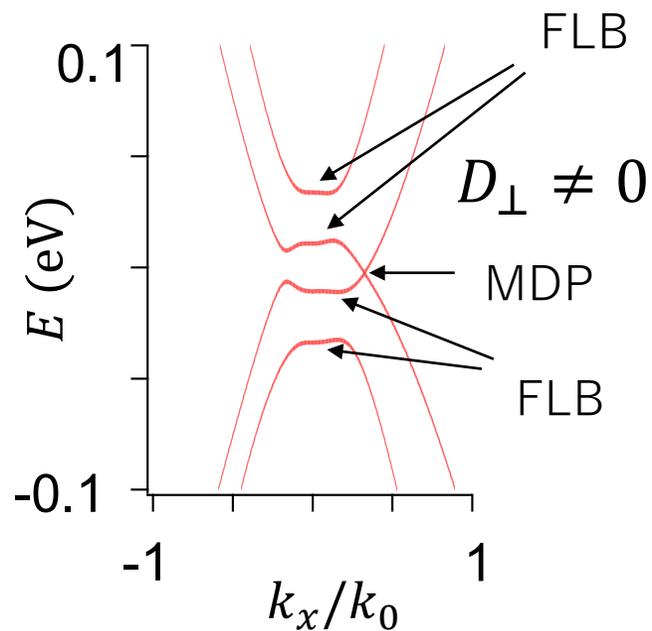
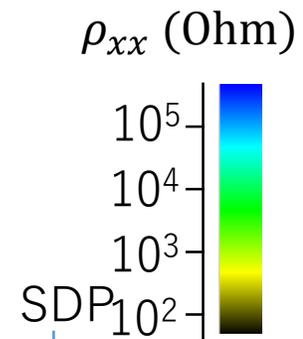
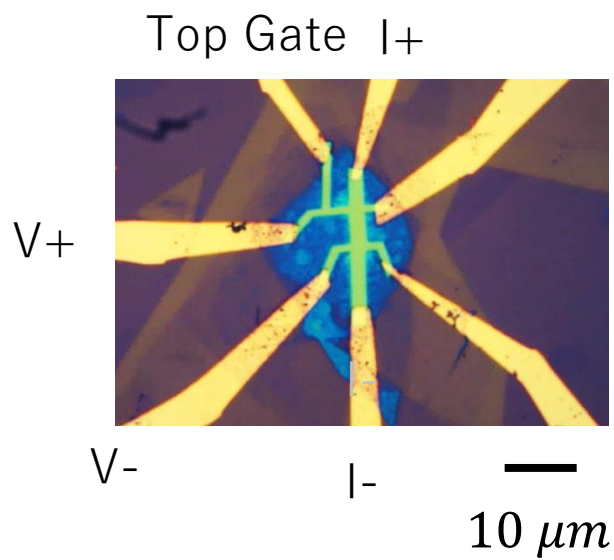
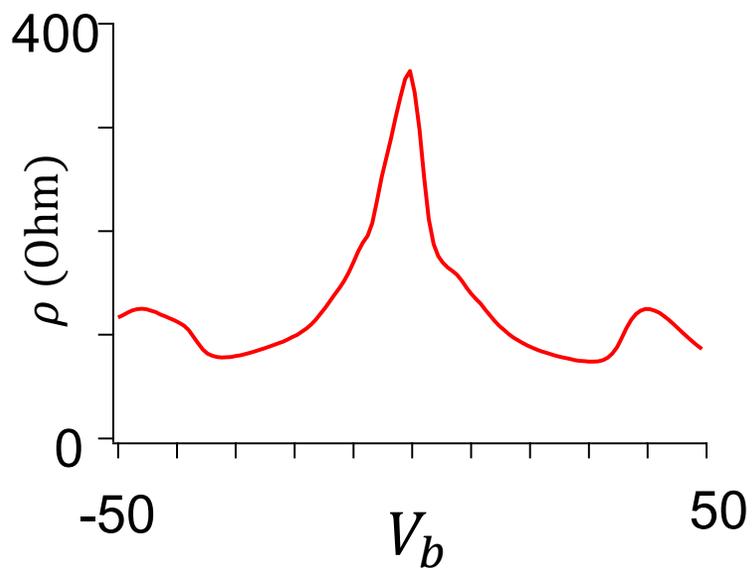
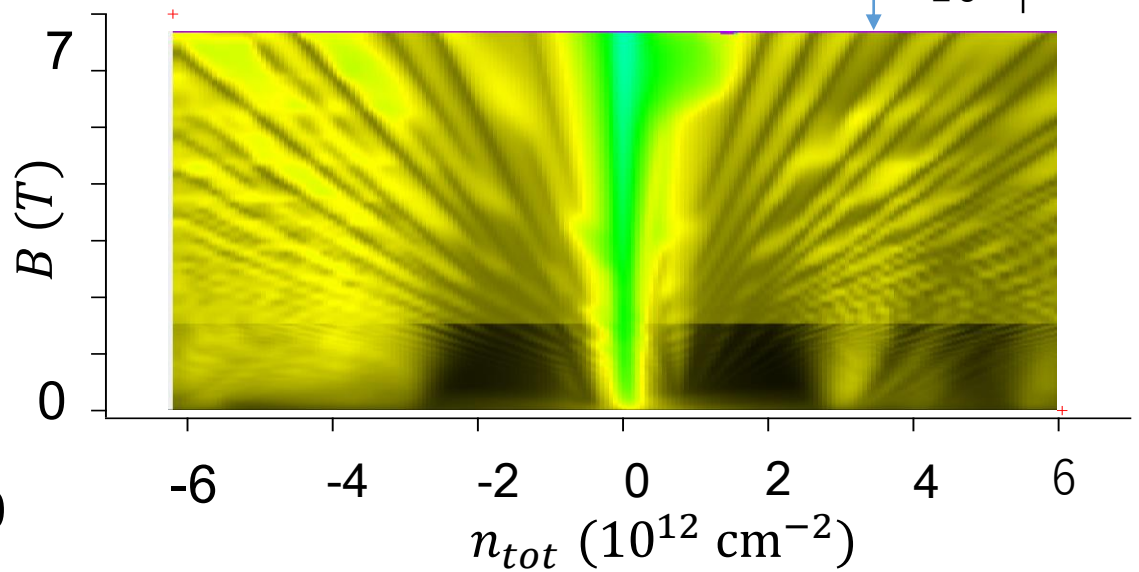

Fig. 1

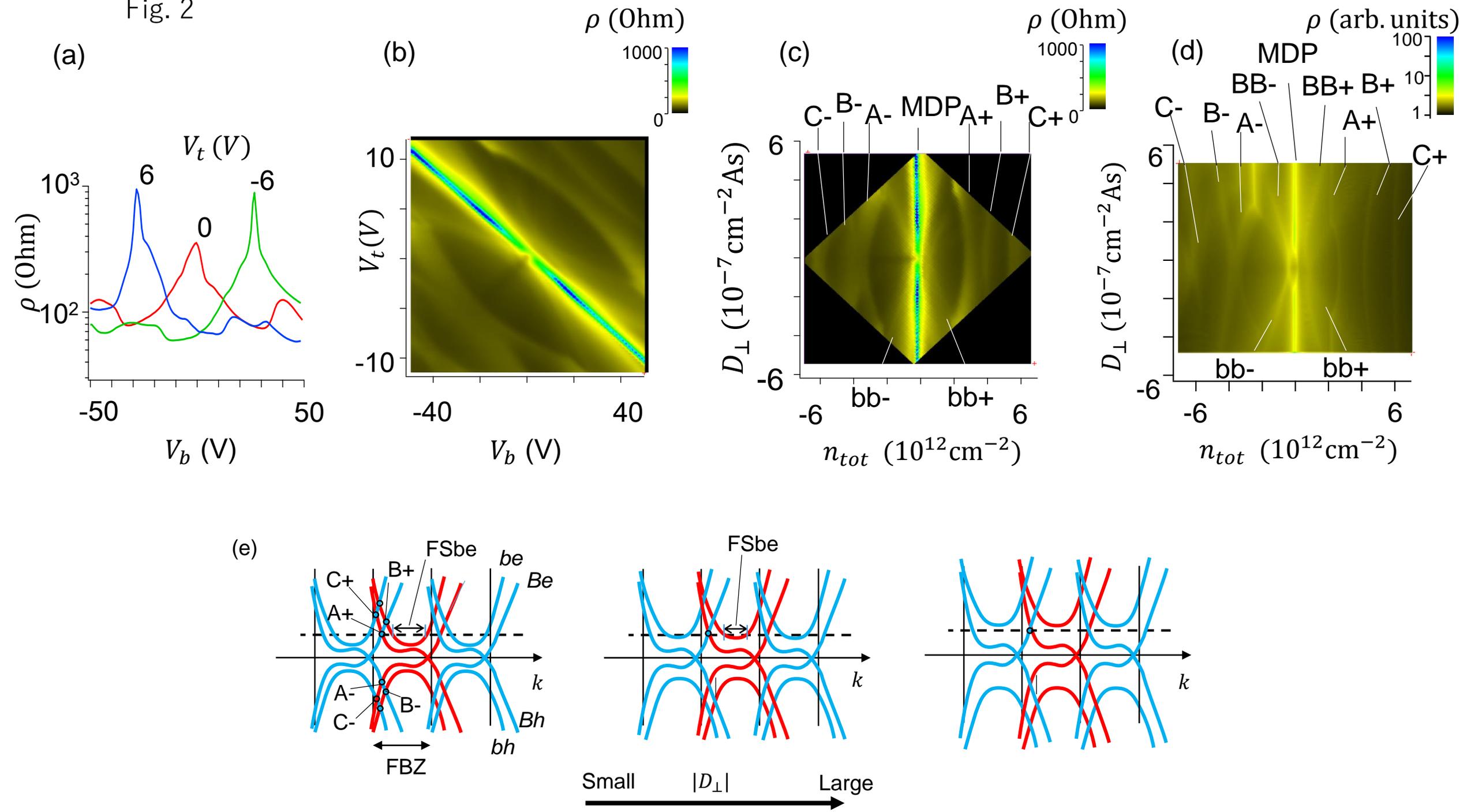

Fig. 3

| | (a) | (b) | (c) |
|---|---|---|---|
| $n_{tot}$ (cm$^{-2}$) | $2.08 \times 10^{12}$ | $4.32 \times 10^{12}$ | $5.95 \times 10^{12}$ |
| $E$ (eV) | 0.0386 | 0.0684 | 0.0894 |

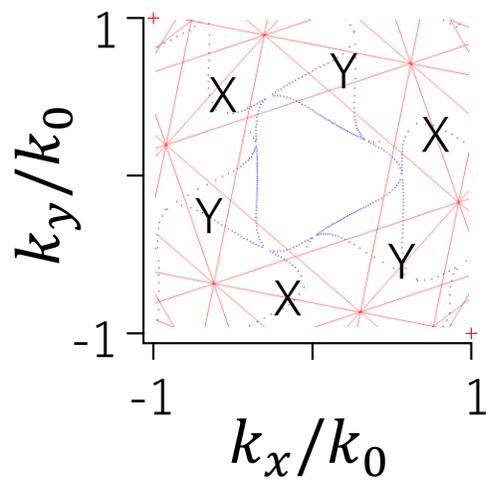
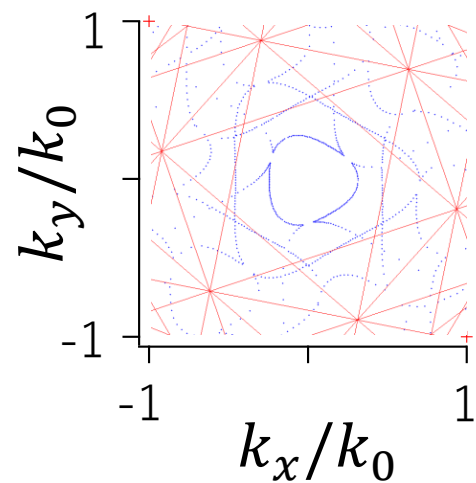
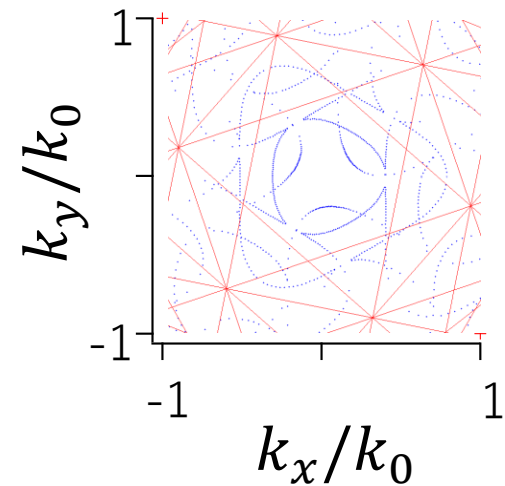
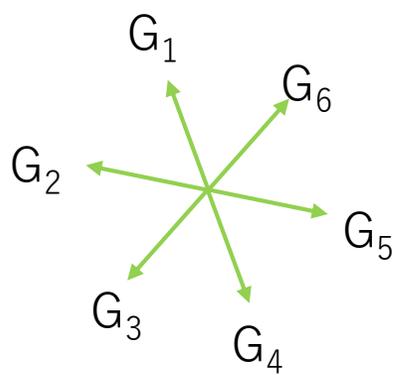
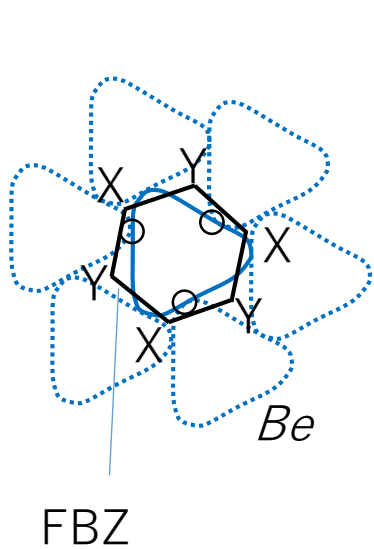
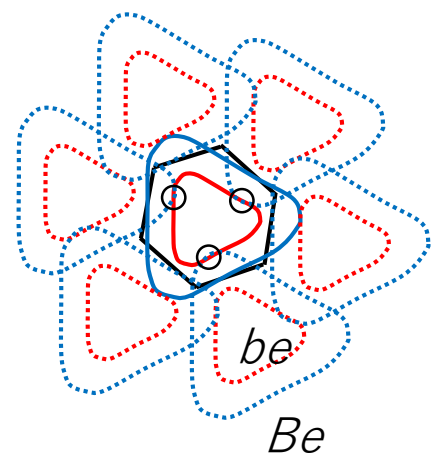
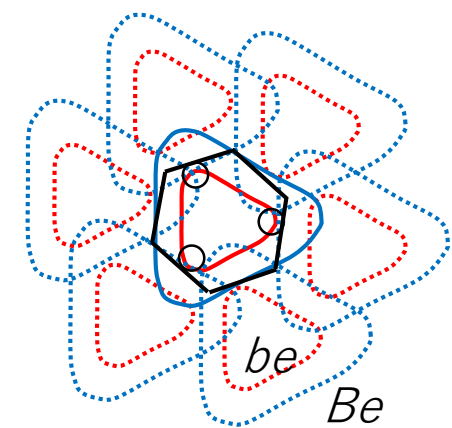

Fig. 4

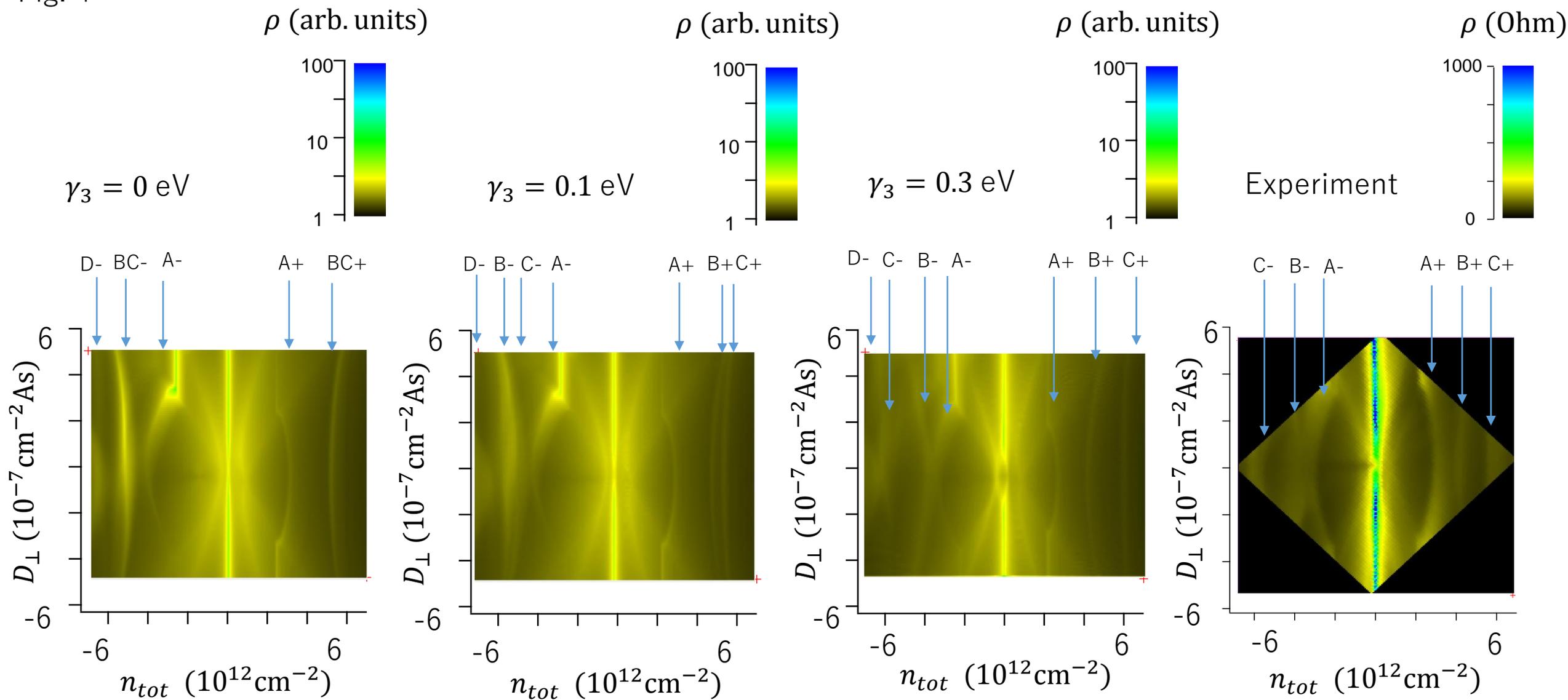

# Supplementary Information

Topological transition of Fermi surface due to nesting in moiré superlattice of *h*-BN/ graphene heterostructure

Fumiya Mukai[1], Kota Horii[1], Ryoya Ebisuoka[2], Kenji Watanabe[2], Takashi Taniguchi[3] and Ryuta Yagi[1*]

[1]Graduate School of Advanced Sciences and Engineering, Hiroshima University,
1-3-1 Kagamiyama, Higashi-Hiroshima 739-8530, Japan,
[2]Research Center for Functional Materials, National Institute for Materials Science, 1-1 Namiki, Tsukuba 305-0044, Japan,
[3]International Center for Materials Nanoarchitectonics, National Institute for Materials Science, 1-1 Namiki, Tsukuba 305-0044, Japan.

## A  Sample fabrication and method of measurement

The schematic structure of our sample is depicted in Figure S1 (a). The sample was *h*-BN-encapsulated AB-stacked tetralayer graphene formed on Si substrate covered with $SiO_2$. The *h*-BN and graphene flakes were prepared by mechanical cleaving using adhesive tape. The stack was formed using dry transfer process with polypropylene carbonate (PPC) film [1]. The sample was equipped with top and bottom gate electrodes. The top one was made of a few-layer graphene and the bottom one was the Si substrate which was heavily doped and remained conducting down to liquid helium temperatures. The sample was patterned into a Hall bar via reactive ion etching using a mixture of low-pressure $CF_4$ and $O_2$ gas. The electric contacts to the graphene were formed by using the edge-contact-technique [1].

The number of graphene layers was roughly determined by making optical contrast measurements [2-4]. Then, the number of layers and stacking were verified via a Landau fan diagram, which is a map of resistivity measured as a function of carrier density and magnetic field. The structure of the Landau levels characterizes the electronic band structures of graphene. It is different for different numbers of layers and stackings, and accordingly, the fan diagrams exhibit specific patterns. By comparing the diagram with those in the past studies [5], the present sample was identified to be AB-stacked tetralayer graphene.

The electrical mobility of the sample was $4 - 6 \times 10^4$ cm$^{-2}$/Vs at low temperature (Fig. S1(b)). The conspicuous decrease in mobility near $|n_{tot}| \leq \pm 2.5 \times 10^{12}$ cm$^{-2}$ is due to the appearance of superlattice peak.

Resistivity measurements were performed by using the standard four-probe method and lock-in technique. The magnetic field was applied using superconducting solenoid.

## B  Effect of perpendicular electric field on moiré superlattice of *h*-BN/bilayer graphene heterostructure.

The dependence of resistivity on the top and bottom gate voltage was studied in a moiré superlattice of a bilayer *h*-BN graphene hetero structure to compare with the case of AB-stacked tetralayer graphene. Pristine bilayer graphene has one massive band at low energy, which contrasts with the AB-stacked tetralayer case with two massive bands. Figure S2 shows a map of resistivity in bilayer graphene as a function of $n_{tot}$ and $D_\perp$. (The sample was patterned into an antidot lattice. However, we think that this is irrelevant to the present discussion because lattice constant of antidot lattice (~ 700 nm) is much larger than the wavelength of the moiré superlattice.) One can see resistance peaks (DP) arising from charge neutrality and superlattice peaks (SP) in Figs. S2 (a) and (b). Peaks near the SP in the hole regime possibly arise from inhomogeneity. The mismatch angle was estimated to be $\theta = 1.11°$ from the carrier density of SP.

Figure S2 (c) is a replot of the map against $n_{tot}$ and $D_\perp$. The structure of the resistance ridges on this map is much simpler than that of the AB-stacked tetralayer case (Fig. 2 (c) in the main text). The resistance ridges appear vertically in the figure, which indicates that $n_{tot}$ for the peaks are approximately unchanged by $|D_\perp|$. The map also shows that the magnitude of the main peak (DP) grows with increasing $|D_\perp|$ because of an energy gap opening. On the other hand, the magnitude of the superlattice peaks were approximately independent of $D_\perp$. This would indicate that the energy gaps associated with these peaks are approximately independent of $D_\perp$.

## C  Method of calculating band structure and resistivity

We calculated the band structure of tetralayer graphene by using the Hamiltonian of the effective mass approximation that considers all the Slonczewski-Weiss-McClure (SWMcC) parameters, *i.e.*, $\gamma_0 - \gamma_5$, and $\delta_p$ [6]. The effective moiré potential in eq. (17)-(18) in Ref. [7] was added to the diagonal element of the first layer of graphene that contacts *h*-BN.

The main text presents the results calculated using the effective moiré Hamiltonian of Moon and Koshino [7], which is given by

$$V_{moiré} = V_0 \begin{pmatrix} 1 & 0 \\ 0 & 1 \end{pmatrix}$$
$$+ \left\{ V_1 e^{i\zeta\psi} \left[ \begin{pmatrix} 1 & \omega^{-\zeta} \\ 1 & \omega^{-\zeta} \end{pmatrix} e^{i\zeta G_1^M \cdot r} + \begin{pmatrix} 1 & \omega^{-\zeta} \\ 1 & \omega^{-\zeta} \end{pmatrix} e^{i\zeta G_1^M \cdot r} + \begin{pmatrix} 1 & \omega^{-\zeta} \\ 1 & \omega^{-\zeta} \end{pmatrix} e^{i\zeta G_1^M \cdot r} \right] + H.C. \right\}.$$

(S1)

Here, $\omega = exp\left(\frac{2\pi}{3}i\right)$, $G_i^M$ ($i = 1, 2, 3$) are the reciprocal lattice vectors of the moiré superlattice, $V_0 = 0.02809$ eV, $V_1 = 0.0210$ eV and $\psi = -0.29$ [7].

On the other hand, Wallbank has derived an effective moiré potential for *h*-BN-graphene heterostructure [8, 9];

$$V_{moiré} = \sum_{m=0}^{5} \left[ U_0^+ + \zeta(-1)^{m+\frac{1}{2}} U_3^+ \sigma_3 - \zeta i (-1)^{m+\frac{1}{2}} U_1^+ \frac{a_m}{a} \cdot \sigma \right] e^{i b_m \cdot r}.$$

(S2)

Here, $\boldsymbol{\sigma} = (\sigma_1, \sigma_2, \sigma_3)$, and $\sigma_i$ ($i = 1, 2, 3$) are the Pauli matrices. $\boldsymbol{a_m}$ is a primitive vector of graphene, which is defined by rotating $(a, 0)$ by $(2\pi m/6)$. $\boldsymbol{b_m}$ is a reciprocal vector of the moiré superlattice. In ref. [9] values of $(U_0^+, U_1^+, U_3^+)$ are given for different models of effective moiré potential. In case of the potential modulation model,

$$(U_0^+, U_1^+, U_3^+) = (V_0, 0, 0) \tag{S3}$$

with $V_0 \sim 60$ meV [9-10]. In the 2D charge modulation model [8, 9, 11], the parameters are given by

$$(U_0^+, U_1^+, U_3^+) = \left(-\frac{V_0}{2}, 0, \frac{\sqrt{3}V_0}{2}\right) \tag{S4}$$

$V_0$ is a phenomenological parameter. We used $V_0 = 60$ meV in the calculation. In the one-site version of the G-hBN hopping model [12], it is approximately given by

$$(U_0^+,\ U_1^+,\ U_3^+) = (V_0/2,\ -V_0,\ -\sqrt{3}\ V_0/2). \tag{S5}$$

The numerical calculation used $V_0 = 30$ meV for a small mismatch angle $\theta$.

To compare with the numerically calculated band structures with the experimental results, we used a potential amplitude scaled by a factor $s$, which served as an adjustable parameter. We also adjusted the mismatch angle $\theta$ between the crystal axes of $h$-BN and graphene.

The dispersion relations were calculated by using the Hamiltonian of the continuum model [6],

$$H = \begin{pmatrix} H_0 + V_{moire} + \Phi_1 & V & W & 0 \\ V^\dagger & H_0' + \Phi_2 & V^\dagger & W' \\ W & V & H_0 + \Phi_3 & V \\ 0 & W' & V^\dagger & H_0' + \Phi_4 \end{pmatrix}. \tag{S6}$$

Here,

$$H_0 = \begin{pmatrix} 0 & v\pi^\dagger \\ v\pi & \Delta' \end{pmatrix},\ H_0' = \begin{pmatrix} \Delta' & v\pi^\dagger \\ v\pi & 0 \end{pmatrix},$$

$$V = \begin{pmatrix} -v_4\pi^\dagger & v_3\pi \\ \gamma_1 & -v_4\pi \end{pmatrix},\ W = \begin{pmatrix} \gamma_2/2 & 0 \\ 0 & \gamma_5/2 \end{pmatrix},\ W' = \begin{pmatrix} \gamma_5/2 & 0 \\ 0 & \gamma_2/2 \end{pmatrix}, \tag{S7}$$

$\pi = \zeta p_x + ip_y$, $v = \sqrt{3}a\gamma_0/2\hbar$, $v_3 = \sqrt{3}a\gamma_3/2\hbar$, and $v_4 = \frac{\sqrt{3}a\gamma_4}{2\hbar}$. Here, $\Phi_i$ is

$$\Phi_i = \begin{pmatrix} -e\phi_i & 0 \\ 0 & -e\phi_i \end{pmatrix}, \tag{S8}$$

and $\phi_i$ ($i = 1 \sim 4$) is the electrostatic potential due to the top and bottom gate voltage. $\phi_i$ reflects the perpendicular electric field so that one must consider screening of the external electric field properly in multilayer graphene [13-16]. We calculated $\phi_i$ by assuming that charges induced by each gate voltage attenuate with a relaxation length of $\lambda \sim 0.45$ nm. [16-19].

Resistivity was numerically calculated by using the constant relaxation time approximation of the Boltzmann transport theory. In this calculation the derivative of the Fermi function with respect to energy was approximated using a Gaussian function with a width of 1 meV.

## D  Maps of resistivity for different effective potential models and Fermi surface topologies

Figures S3-S6 show maps of resistivity numerically calculated for different models of effective moiré potential. Figure S3 is for the *h*-BN-graphene hopping model of Moon and Koshino [7], Fig. S4 is for the one-site version of the G-hBN hopping model [9,12], S5 is for the 2D charge modulation model [9,11], and S6 is for the potential modulation model[10]. In each figure, the factor $s$ was varied from left to right as 0.25, 0.5 and 1, with the right-most panel showing the experimental results. It can be seen that the resistance ridge structures are strongly dependent on *s*, which reflects the variation in the detailed structure in the dispersion relations. The maps are slightly asymmetric in $D_\perp = 0$. This is because the effective moiré potential is present only in the first layer of graphene in the Hamiltonian, which results in an effective offset in $D_\perp$. In all the potential models, the map for $s = 1$ shows a significant difference from the experiment. For example, in Fig. S3, a significant difference can be seen in resistance ridges A+, A- ,bb+, and bb-; ridges A+ and A- are curved more tightly at small values of $|D_\perp|$, and show kinks before they merge and straighten out for larger $D_\perp$. Also, $n_{tot}$ for point G, where bb+ and BB+ crosses at $D_\perp = 0$, is about $1.2 \times 10^{12}$ cm$^{-2}$ in the calculation while $n_{tot} \approx 0.8 \times 10^{12}$ cm$^{-2}$ in the experiment. This would indicate that the effective potential is too large to reproduce the experimental values. All potential models approximately reproduced the experimental results when $s \leq 0.5$.

The fact that the patterns of the resistance ridges are commonly reproduced for a small effective moiré potential would mean that the ridge structure is principally determined by the geometry of the superlattice Brillouin zone and the topology of the Fermi surface. This is verified in Figs S8-S10 which show the energy contours of the band structures for effective moiré potential models with $s = 0.5$. The different effective moiré potential results in Fermi surfaces with slightly different shapes. However, in all of these figures, the Fermi surfaces (lines) disappear when the Fermi surfaces in different Brillouin zone make contact (nests) in the extended zone scheme, which results in a topological transition. Because the Fermi surface of multilayer graphene is significantly trigonally warped, an energy gap does not necessarily open at the zone boundary. This contrasts with the case of monolayer graphene, a system with an approximately isotropic Fermi surface, where the energy gap opens approximately on the boundary of the superlattice Brillouin zone.

### F  Determining mismatch angle between *h*-BN and graphene

For monolayer graphene, one can use eqs. (3) and (4) in the main text to determine the mismatch angle and the superlattice size, from the carrier density of the secondary Dirac point. The equations could possibly be applied in the case of bilayer graphene. However, for multiband systems, such as AB-stacked tetralayer graphene, they cannot be used for two reasons. The one is that carriers are distributed in two bands. The other is that the peaks would appear before the Brillouin zone fills up because of the trigonally warped Fermi surface. For these reasons, we determined the mismatch angle by calculating the maps of resistivity for different mismatch angles and compared the map with the experiment. Figure S11 shows maps of resistivity calculated for different mismatch angles. As can be seen in the figure, the experimental map were approximately reproduced for $\theta = 0.35°$.

Figure Captions

Fig. S1   Sample profile
(a) Schematic drawing of the sample structure. TG means top gate, BG, bottom gate. (b) Electric mobility ($\mu$) *vs* $n_{tot}$, measured at $V_t = 0$ V, $T = 4.2$ K and $B = 0$ T.

Fig. S2   Top and bottom gate voltage dependence of resistivity in moiré superlattice of *h*-BN/ bilayer graphene heterostructure
(a) Map of $\rho$ *vs* $V_b$ and $V_t$. $T = 4.2$ K and $B = 0$ T. DP indicates resistance peaks for charge neutrality. SPs are those for the secondary Dirac point. (b) $\rho$ *vs* $V_b$ measured at $V_t = 0$ V. (c) Replot of the map as a function of $n_{tot}$ and $D_\perp$.

Fig. S3. Numerically calculated maps of resistivity I
Numerically calculated maps of resistivity by using the effective moiré potential of Moon and Koshino[7]. *s* is the scaling factor of the potential. $\theta = 0.35°$. The SWMcC parameters of graphite were used in the calculation.

Fig. S4. Numerically calculated maps of resistivity II
Numerically calculated maps of resistivity by using the effective moiré potential of the on-site version of the G-hBN hopping model [8, 9,12]. *s* is the scaling factor of the potential. $\theta = 0.35°$. The SWMcC parameters of graphite were used in the calculation.

Fig. S5 Numerically calculated maps of resistivity III
Numerically calculated maps of resistivity by using the effective moiré potential of the 2D charge modulation model [8, 9,11]. *s* is the scaling factor of the potential. $\theta = 0.35°$. The SWMcC parameters of graphite were used in the calculation.

Fig. S6 Numerically calculated maps of resistivity IV
Numerically calculated maps of resistivity by using the effective moiré potential of the potential modulation model [8-10]. *s* is the scaling factor of the potential. The SWMcC parameters of graphite were used in the calculation.

**Fig. S7 Energy contours of dispersion relation I.**

Numerically calculated energy contours of the dispersion relations by using the effective moire potential of Moon and Koshino[7]. $s$ is the scaling factor of the potential. Here, $\theta = 0.35°$, $D_\perp = 0$, $s = 0.5$, and the SWMcC parameters of graphite were used in the calculation. $k_0 = \frac{2\gamma_1}{\sqrt{3}\gamma_0 a}$ ($a$ is the lattice constant of graphene)

**Fig. S8 Energy contours of dispersion relations II.**

Numerically calculated energy contours of the dispersion relations by using the on-site version of the G-hBN hopping model [9,12]. Here, $\theta = 0.35°$, $D_\perp = 0$, $s = 0.5$, and the SWMcC parameters of graphite were used in the calculation. $k_0 = \frac{2\gamma_1}{\sqrt{3}\gamma_0 a}$ ($a$ is the lattice constant of graphene)

**Fig. S9 Energy contours of dispersion relations III.**

Numerically calculated energy contours of the dispersion relations by using the 2D charge modulation model [9,11]. Here, $\theta = 0.35°$, $D_\perp = 0$, $s = 0.5$, and the SWMcC parameters of graphite were used in the calculation. $k_0 = \frac{2\gamma_1}{\sqrt{3}\gamma_0 a}$ ($a$ is the lattice constant of graphene)

**Fig. S10 Energy contours of dispersion relations IV.**

Numerically calculated energy contours of the dispersion relations by using the potential modulation model [9-10]. Here, $\theta = 0.35°$, $D_\perp = 0$, $s = 0.5$, and the SWMcC parameters of graphite were used in the calculation. $k_0 = \frac{2\gamma_1}{\sqrt{3}\gamma_0 a}$ ($a$ is the lattice constant of graphene)

**Fig. S11 Determination of the mismatch angle**

Maps of resistivity numerically calculated for different mismatch angles, $\theta = 0.2°$, $0.35°$, and $0.5°$. The effective moiré potential of Moon and Koshino [7] was used. The experimental results were approximately reproduced for $\theta = 0.35°$.

# References

[1] Wang, L. *et al.*, One-dimensional electrical contact to a two-dimensional material, Science **342**, 614-617 (2013).

[2] Blake, P. *et al.*, Making graphene visible, Appl. Phys. Lett. **91**, 063124 (2007).

[3] Teo, G. Q. *et al.*, Visibility study of graphene multilayer structures, J. Appl. Phys. **103**, 124302 (2008).

[4] Abergel, D.S.L., Russell, A. & Fal'ko, V. I., Visibility of graphene flakes on a dielectric substrate, Appl. Phys. Lett. **91**, 063125 (2007).

[5] Yagi, R. *et al.*, Low-energy band structure and even-odd layer number effect in AB-stacked multilayer graphene, Sci. Rep. **8**, 13018 (2018).

[6] Koshino, M. & McCann, E., Landau level spectra and the quantum Hall effect of multilayer graphene, Phys. Rev. B **83**, 165443 (2011).

[7] Moon, P. & Koshino, M., Electronic properties of graphene/hexagonal-boron-nitride moire superlattice, Phys. Rev. B **90**, 155406 (2014).

[8] Wallbank, J., Mucha-Kruczynski, M., Chen, X. & Fal'ko, V., Moire superlattice effects in graphene/boron-nitride van der Waals heterostructures, ANN. PHYS. **527**, 359 - 376 (2015).

[9] Wallbank, J., Patel, A., Mucha-Kruczynski, M., Geim, A. & Falko, V., Generic miniband structure of graphene on a hexagonal substrate, Phys. Rev. B **87**, 245408 (2013).

[10] Yankowitz, M. *et al.*, Emergence of superlattice Dirac points in graphene on hexagonal boron nitride, Nat. Phys. **8**, 382 - 386 (2012).

[11] Ortix, C., Yang, L. & van den Brink, J., Graphene on incommensurate substrates, Phys. Rev. B **86**, 081405 (2012).

[12] Kindermann, M., Uchoa, B. & Miller, D., Zero-energy modes and gate-tunable gap in graphene on hexagonal boron nitride, Phys. Rev. B **86**, 115415 (2012).

[13] Guinea, F., Charge distribution and screening in layered graphene systems, Phys. Rev. B **75**, 235433 (2007).

[14] Visscher, P. B. & Falicov, L. M., Dielectric screening in a layered electron gas, Phys. Rev. B **3**, 2541 - 2547 (1971).

[15] Miyazaki, H. *et al.*, Inter-layer screening length to electric field in thin graphite film, Appl. Phys. Express **1**, 034007 (2008).

[16] Koshino, M., Interlayer screening effect in graphene multilayers with ABA and ABC Stacking, Phys. Rev. B **81**, 125304 (2010).

[17] Hirahara, T. *et al.*, Multilayer Graphene Shows Intrinsic Resistance Peaks in The Carrier Density Dependence, Sci. Rep. **8**, 13992 (2018).

[18] Nakasuga, T. *et al.*, Low-energy Band Structure in Bernal Stacked Six-layer

Graphene, Phys. Rev. B **99**, 085404 (2019).

[19] Nakasuga, T. *et al.*, Intrinsic resistance peaks in AB-stacked multilayer graphene with odd number of layers, Phys. Rev. B **101**, 035419 (2020).

Fig. S1

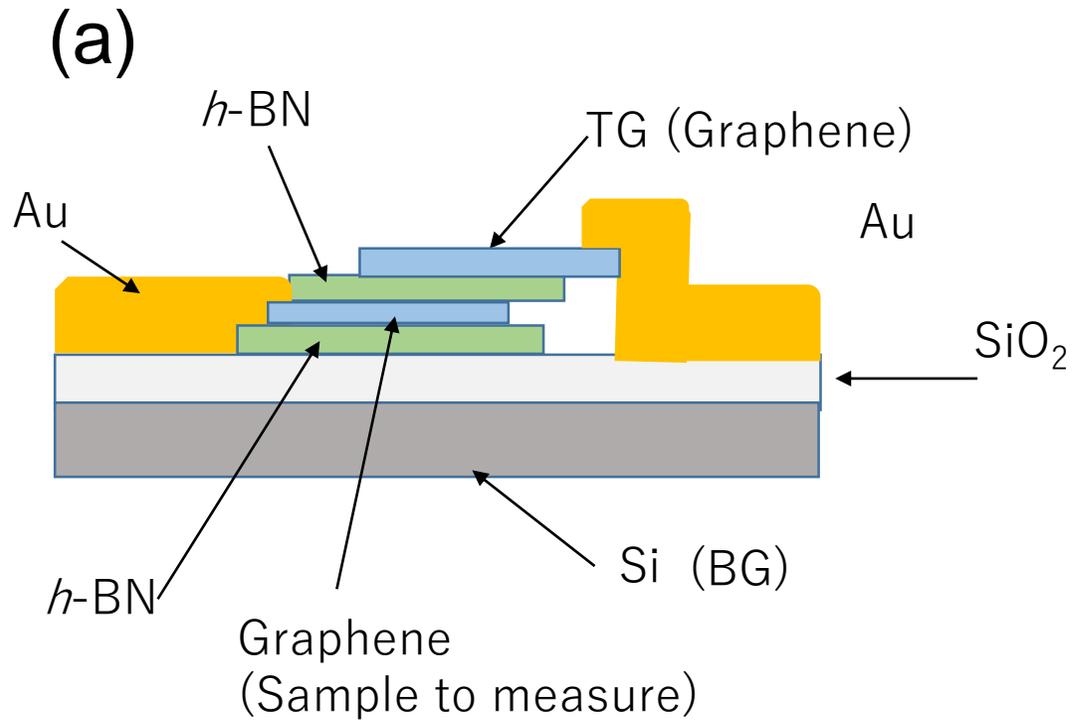 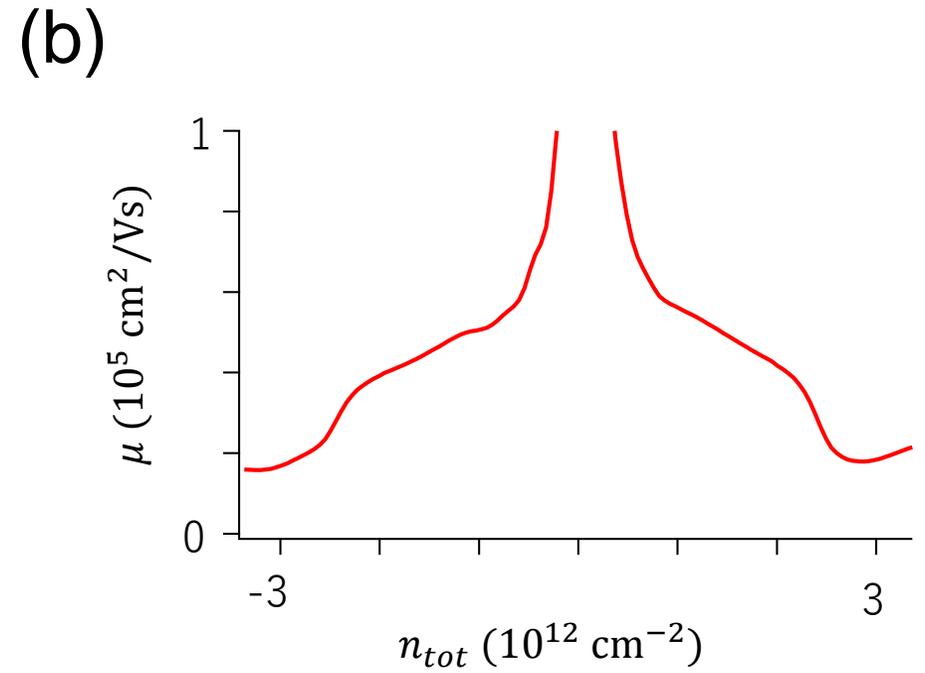

Fig. S2

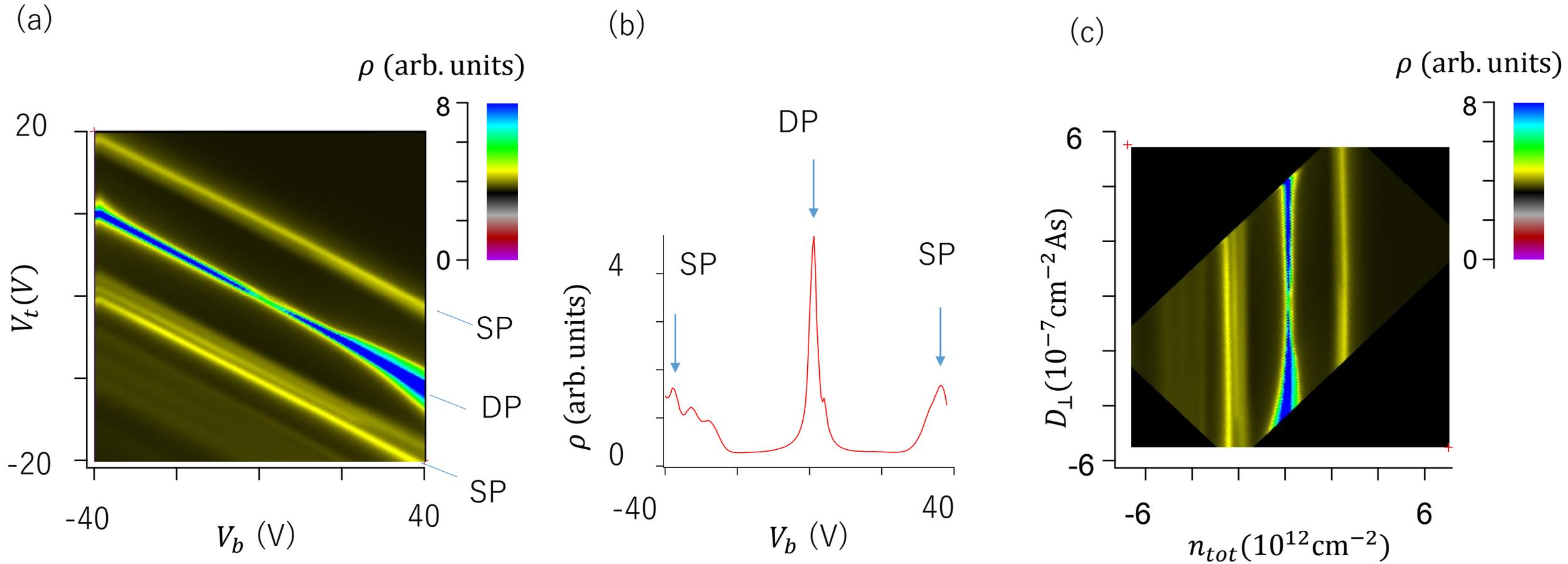

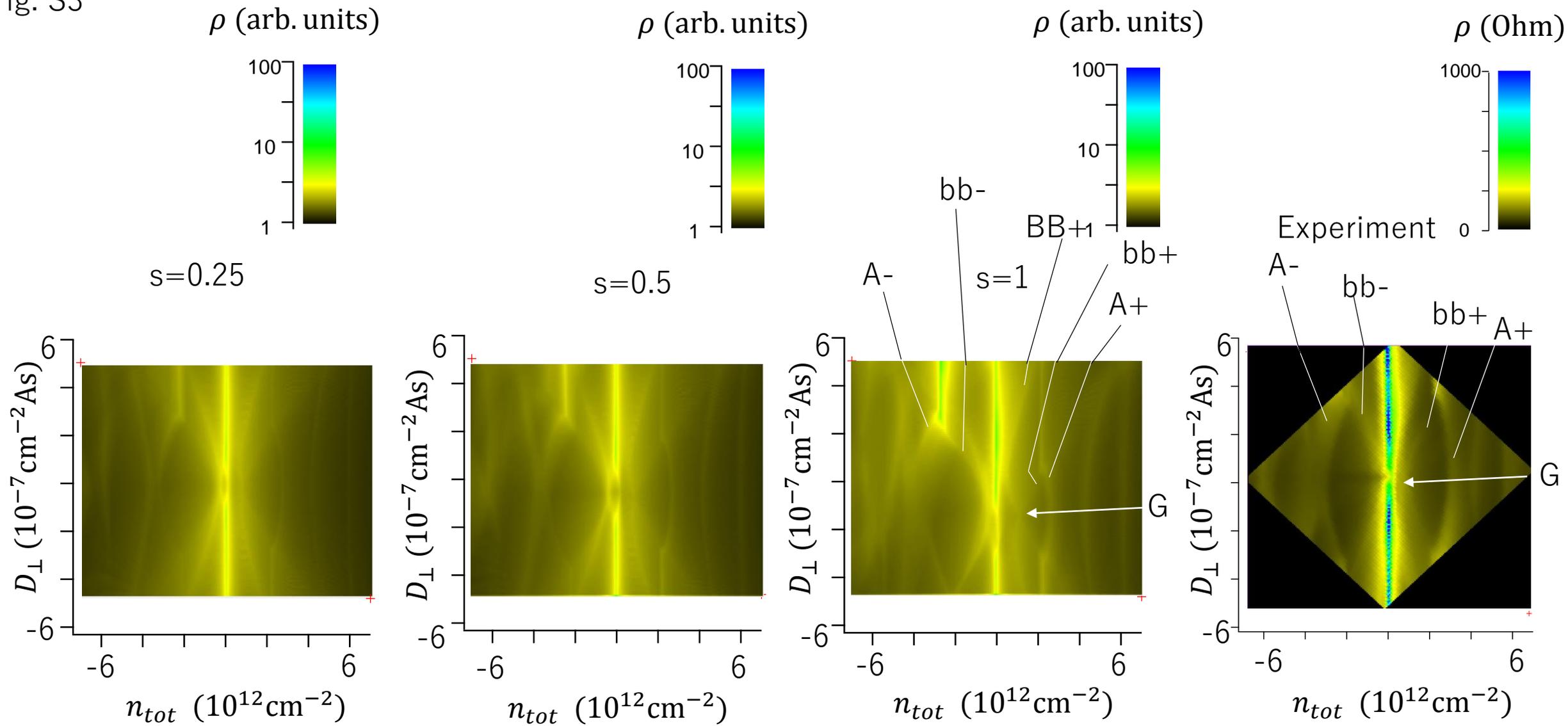

Fig. S3

Fig. S4

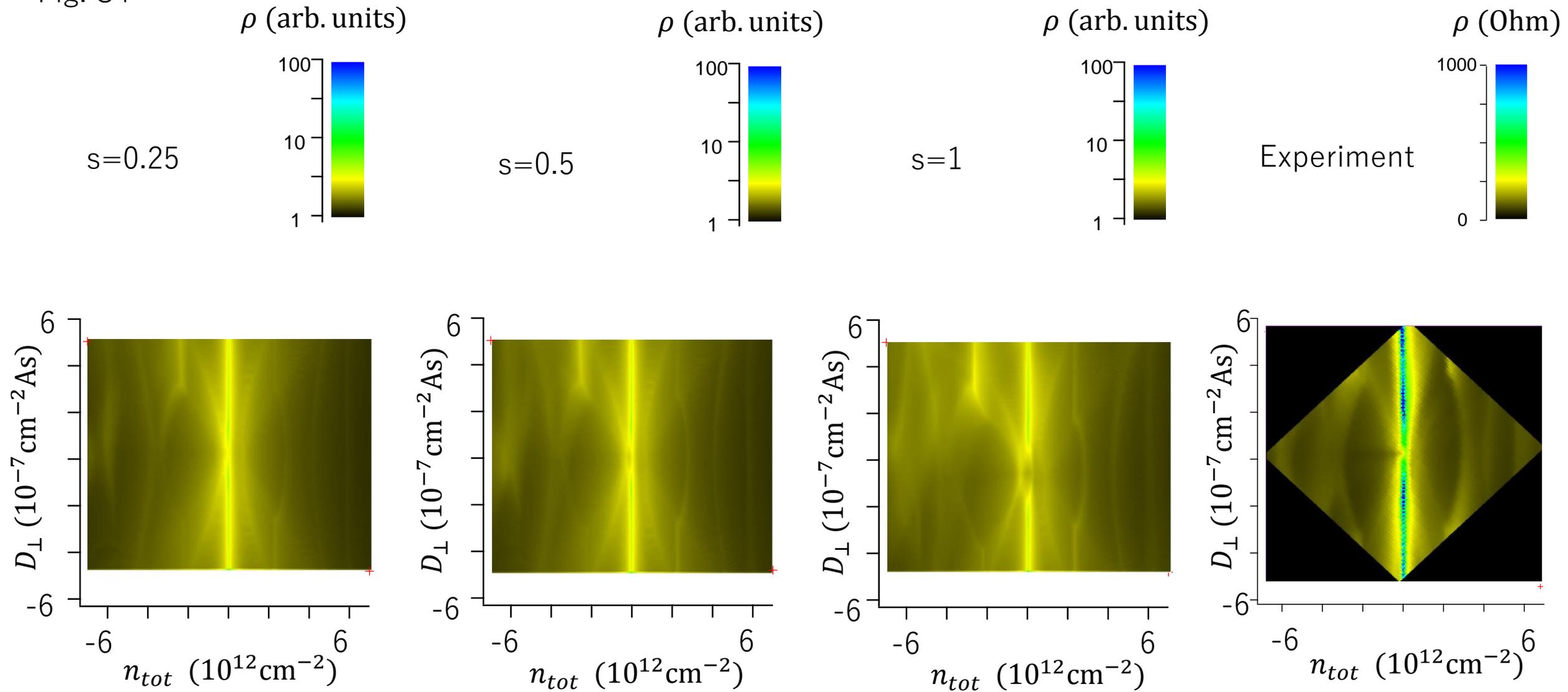

Fig. S5

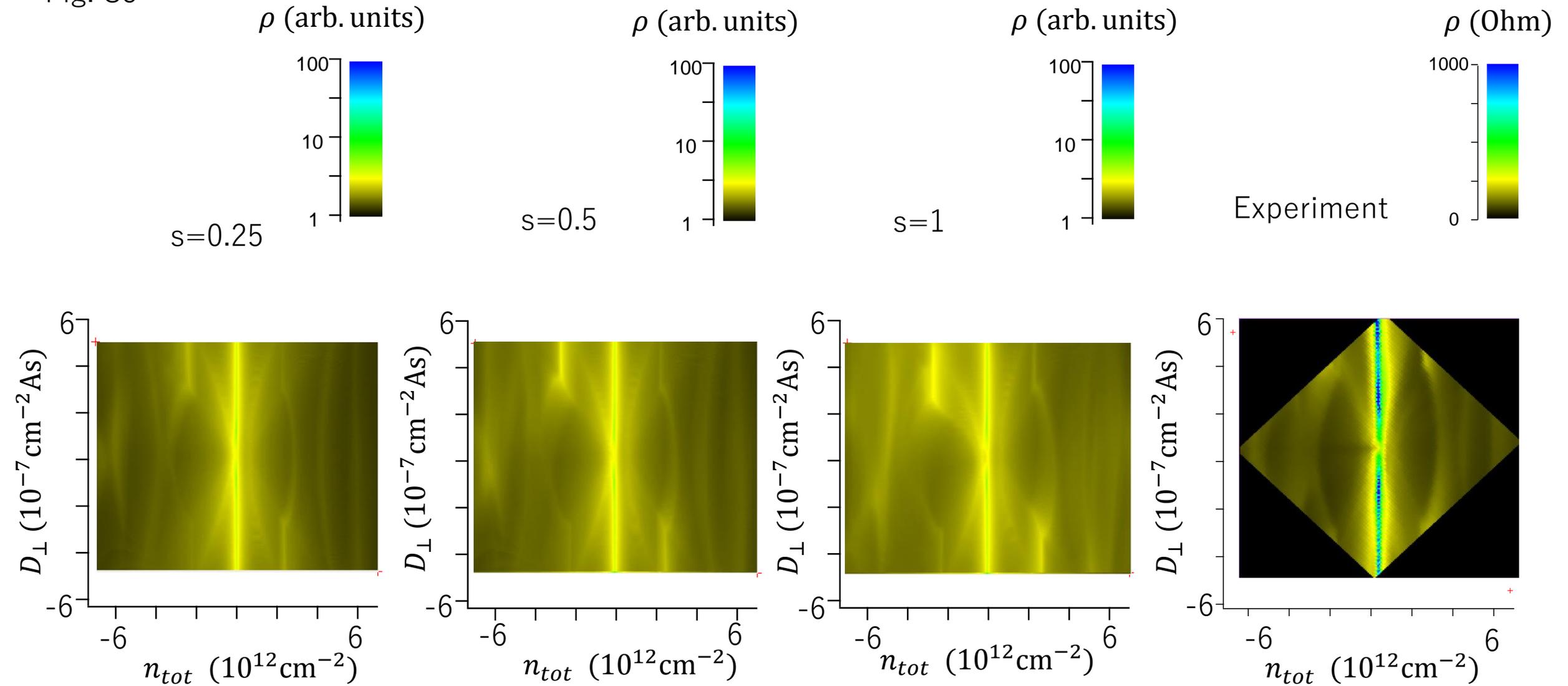

Fig. S6

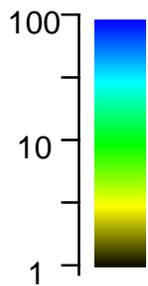 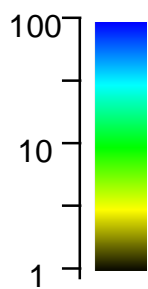 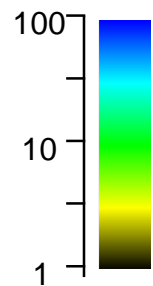 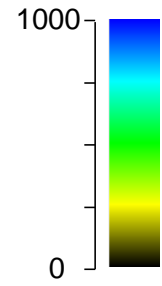

s=0.25      s=0.5      s=1      Experiment

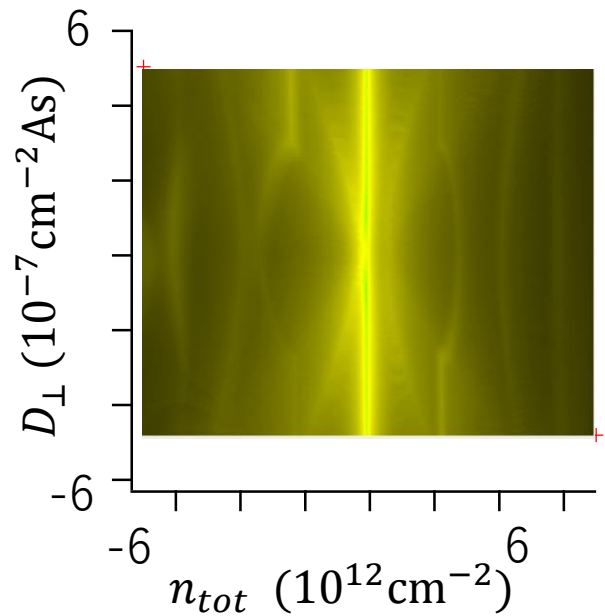 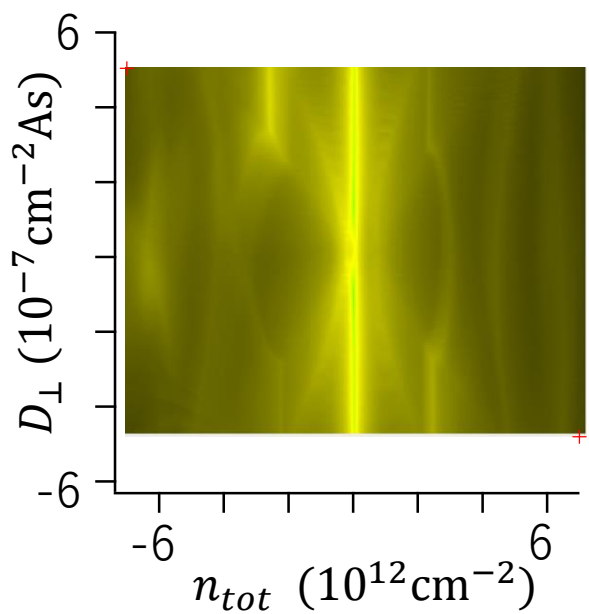 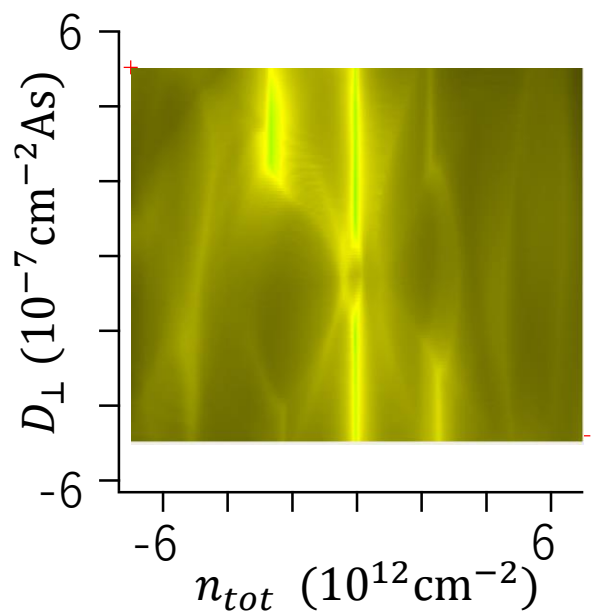 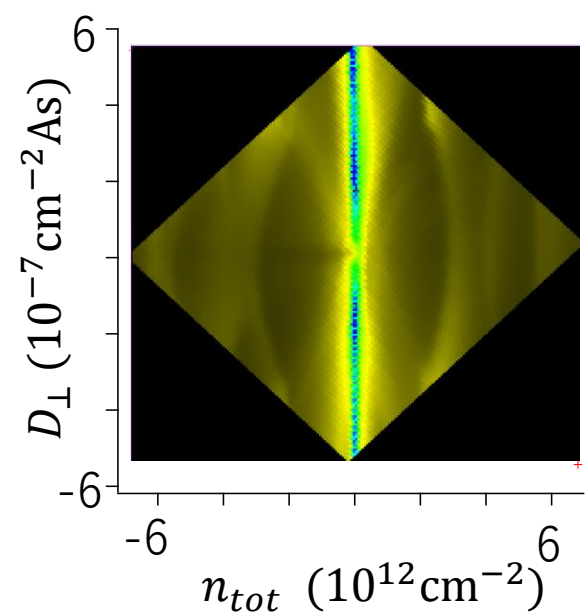

Fig. S7

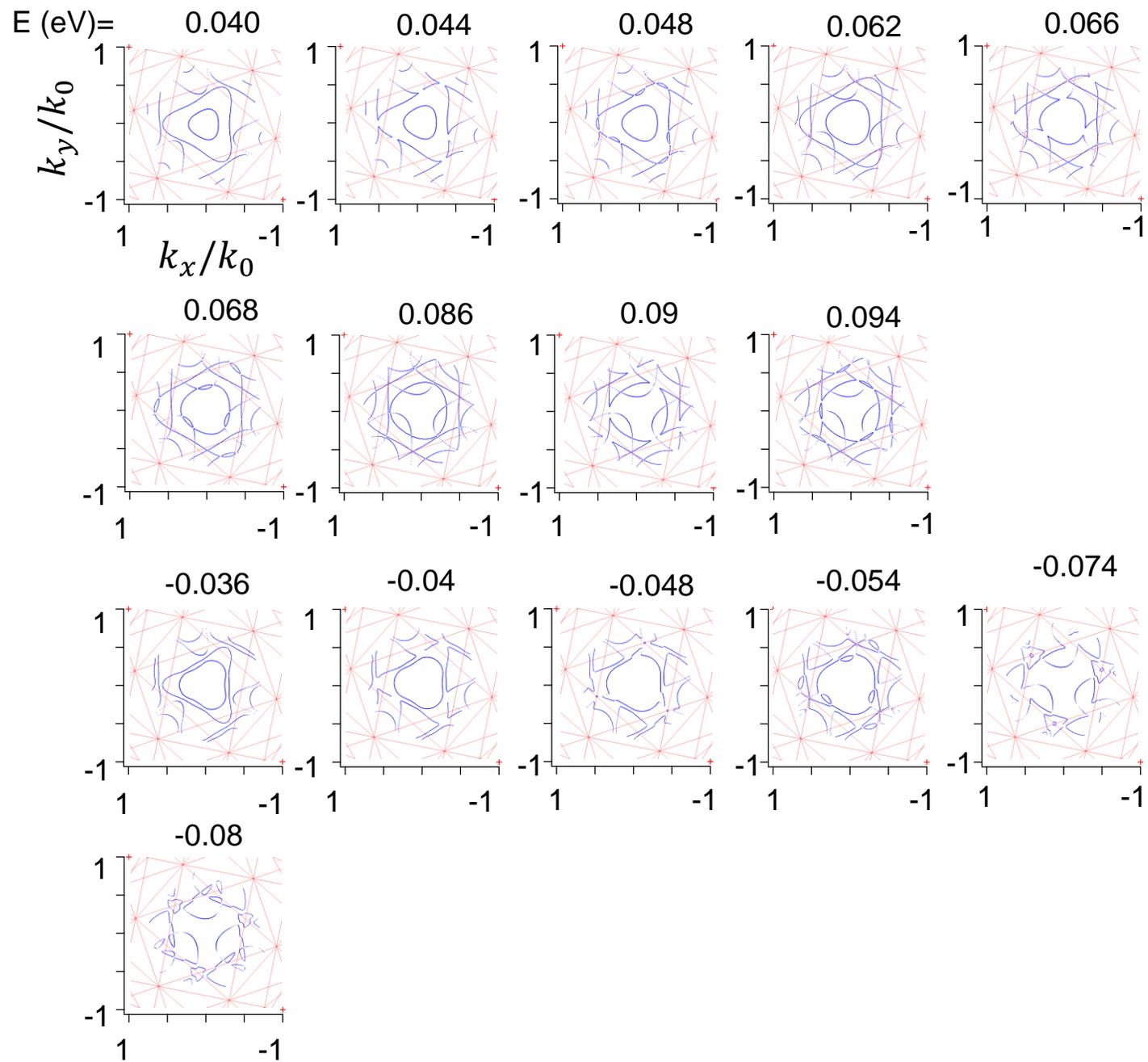

Fig. S8

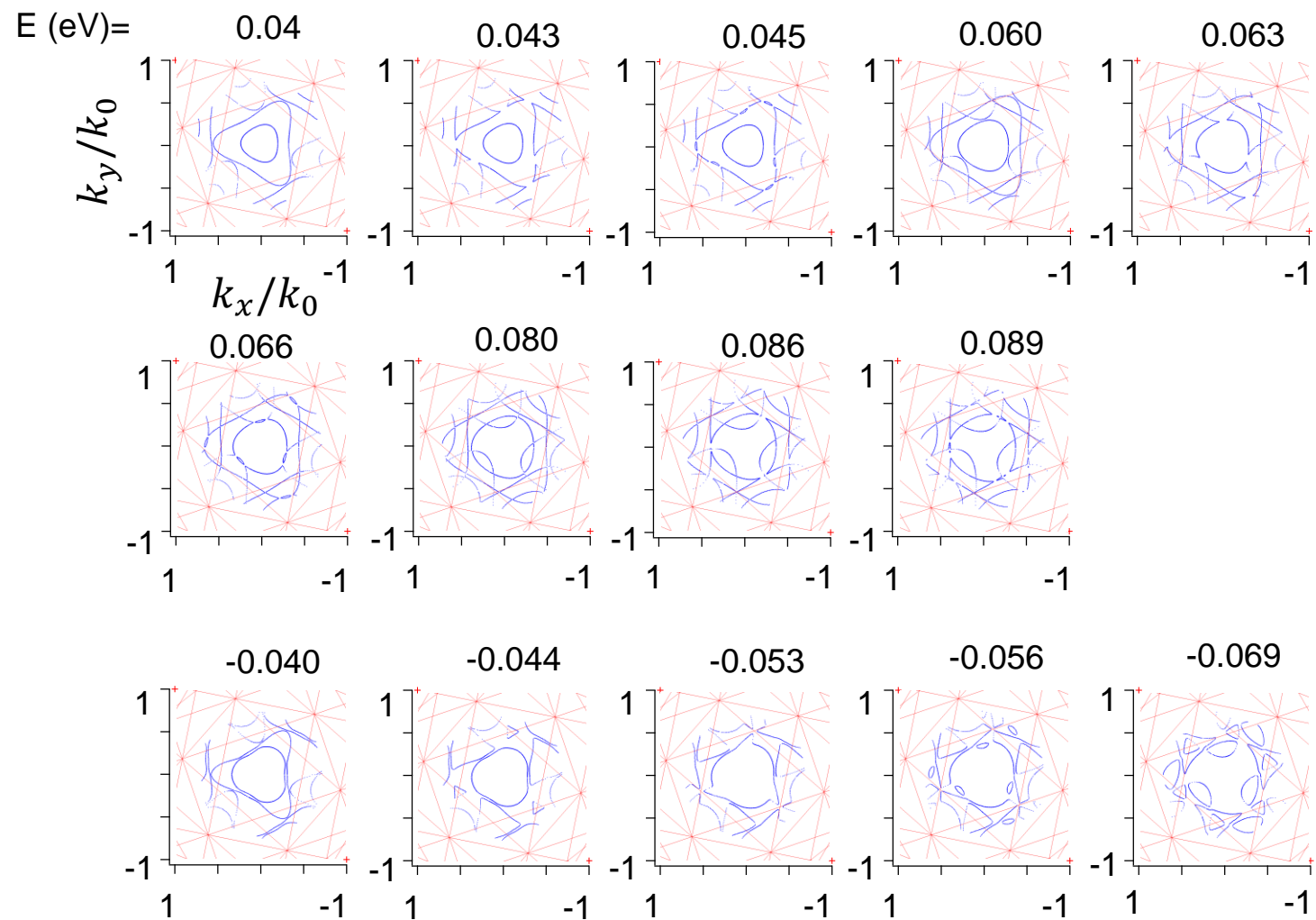

Fig. S9

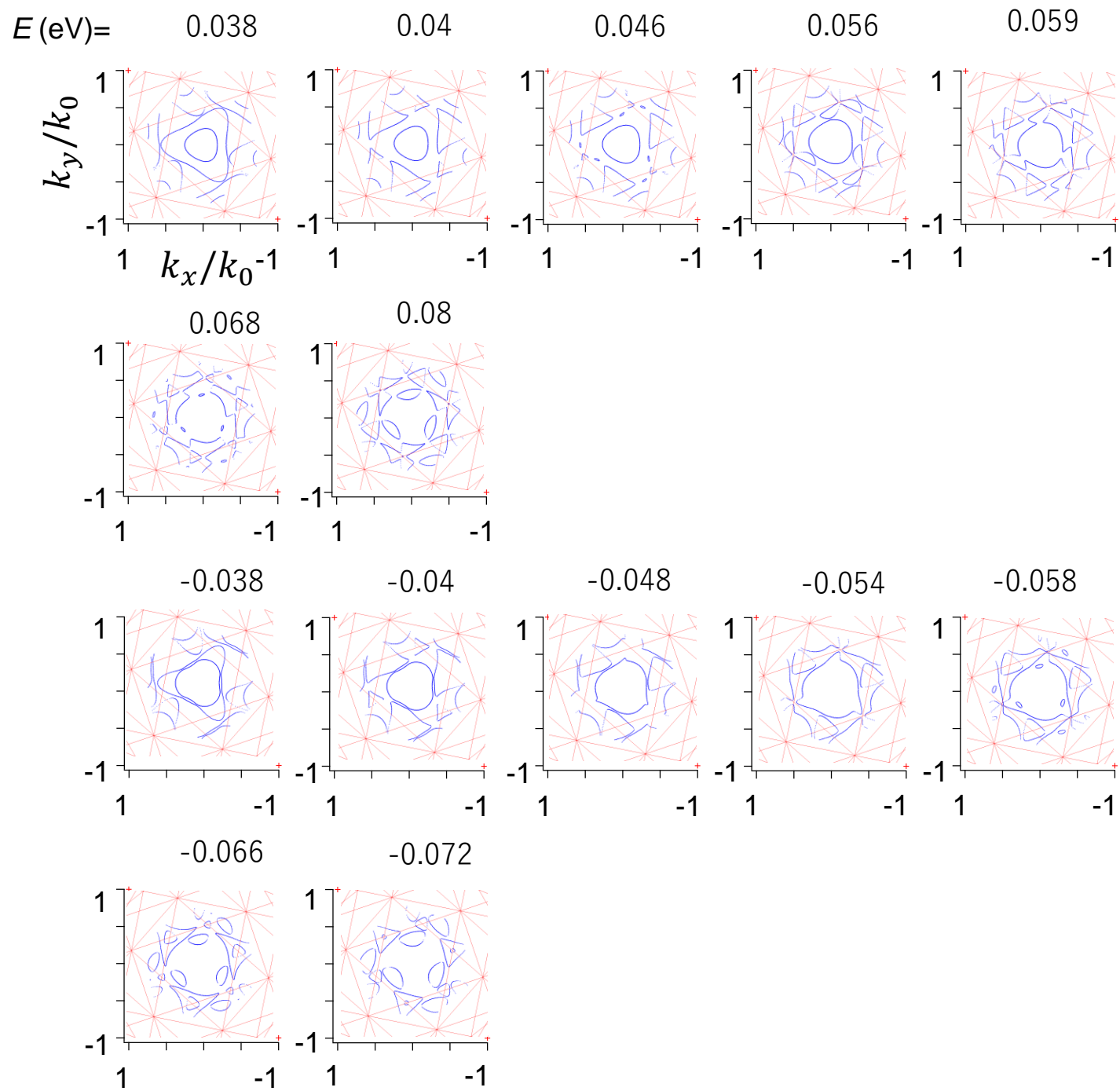

Fig. S10

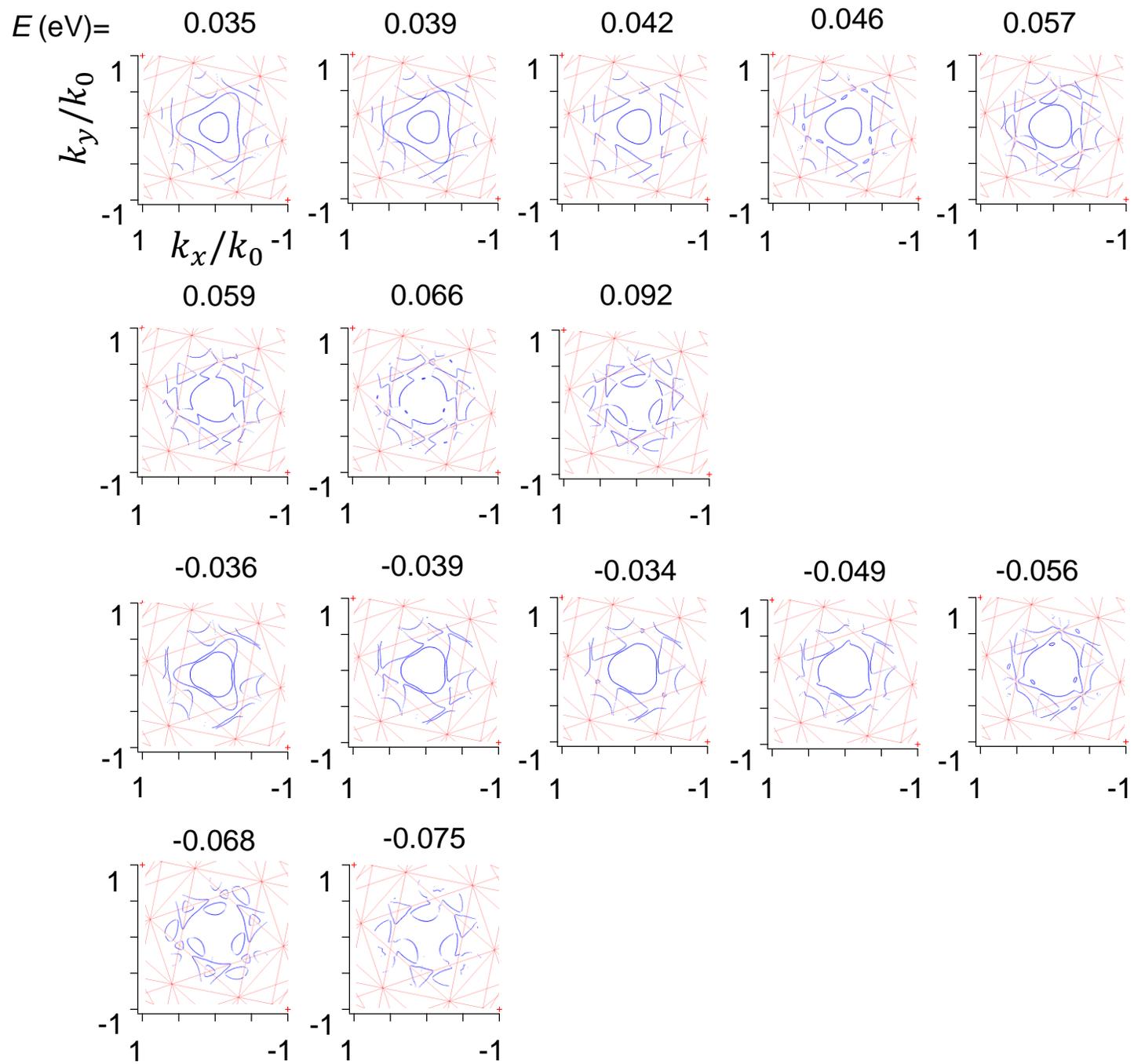

Fig. S11

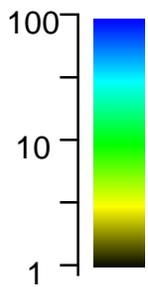 ρ (arb. units)

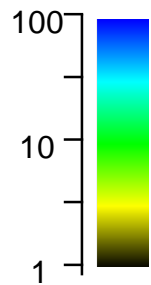 ρ (arb. units)

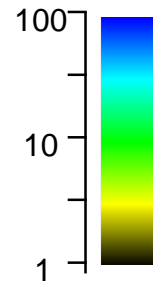 ρ (arb. units)

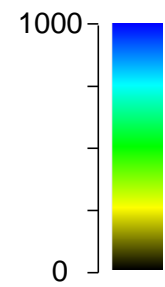 ρ (Ohm)

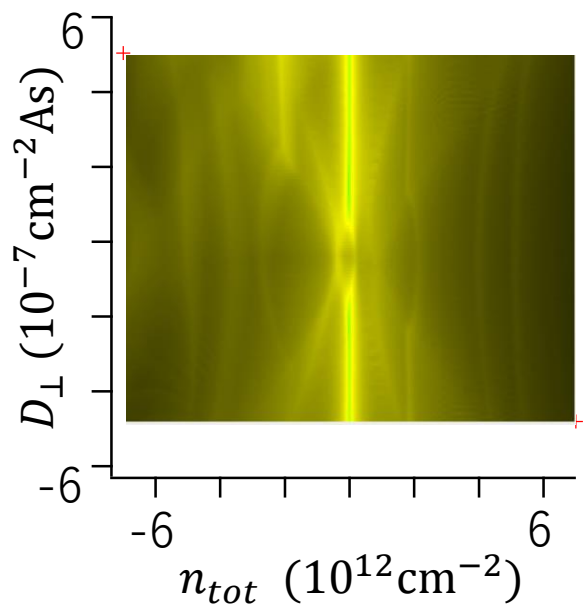 $\theta = 0.2°$

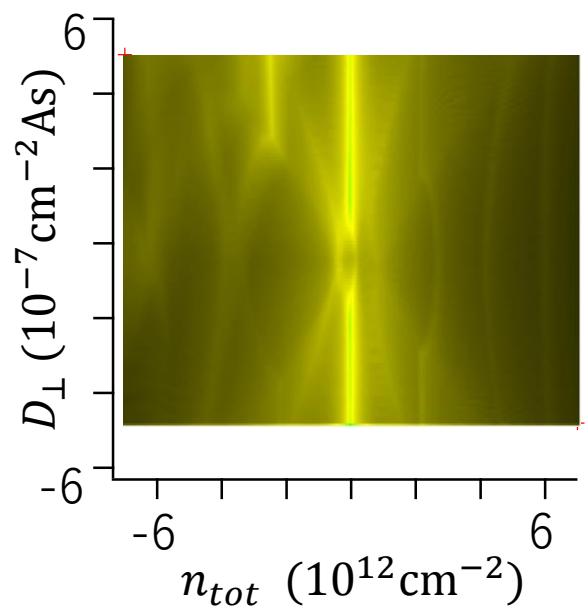 $\theta = 0.35°$

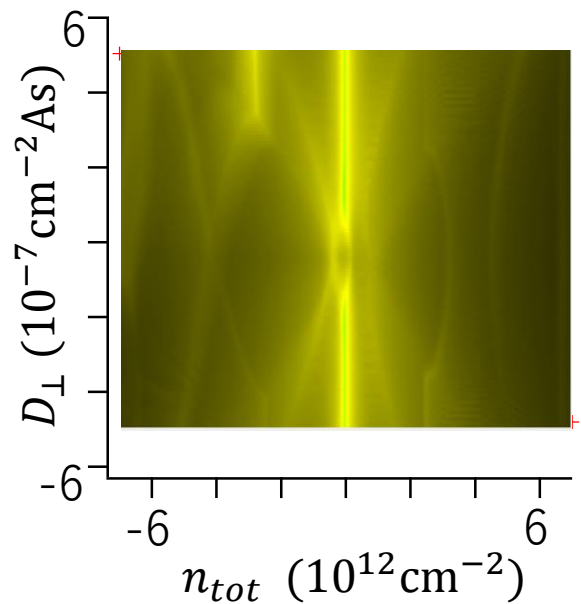 $\theta = 0.5°$

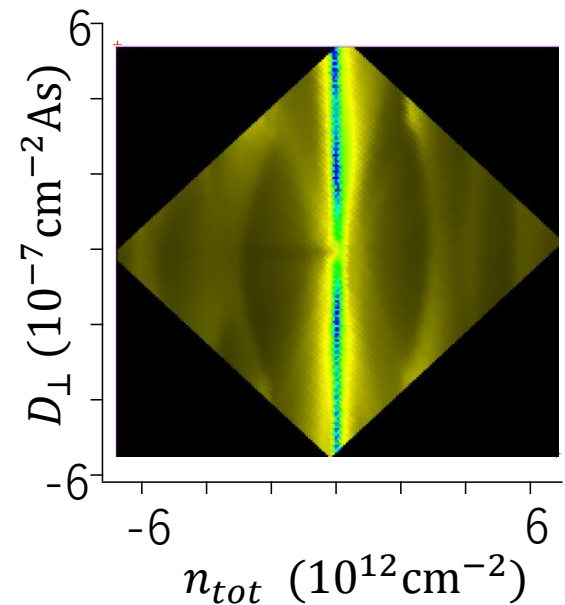 Experiment